\documentclass{aa}  

\usepackage{graphicx}
\usepackage{txfonts}
\usepackage[colorlinks=true,allcolors=blue]{hyperref}
\usepackage{xcolor}
\usepackage{orcidlink}

\definecolor{apcolor}{HTML}{b3003b}

\definecolor{anna}{rgb}{0.00, 0.5, 0.8}

\def\msun{{\rm M}_{\odot}}
\def\msunyr{ {\rm M}_{\odot}/{\rm yr}}
\def\cc{{\rm cm}^{-3}}

\begin{document}
   \title{
   Time evolution of the galactic $\rm {B-\rho}$ relation:\\ the impact of the magnetic field morphology
   }

\authorrunning{Konstantinou et al.}
\titlerunning{The $B-\rho$ relation in galaxies}

   \subtitle{}

   \author{A. Konstantinou
           \inst{1,2 \orcidlink{0000-0002-4758-212X}}\fnmsep\thanks{akonstantinou@physics.uoc.gr}, 
           E. Ntormousi
           \inst{3,2 \orcidlink{0000-0002-4324-0034}}\fnmsep\thanks{evangelia.ntormousi@sns.it}, 
           K. Tassis\inst{1,2 \orcidlink{0000-0002-8831-2038}}
           \and
           A. Pallottini\inst{3 \orcidlink{0000-0002-7129-5761}}
          }

   \institute{Department of Physics, University of Crete, Voutes, 70013 Heraklion, Greece
         \and
             Institute of Astrophysics, Foundation for Research and Technology-Hellas, Voutes, 70013 Heraklion, Greece
         \and Scuola Normale Superiore, Piazza dei Cavalieri, 7 56126 Pisa, Italy
             }

   \date{Received ; accepted }

 
\abstract
  {
  One of the most frequently used indicators to characterize the magnetic field's influence on star formation is the relation between magnetic field strength and gas density ($\rm {B-\rho}$ relation), usually expressed as a power law of the form $\rm B\propto\rho^{\kappa}$. The value of $\kappa$ is an indication of the dynamical importance of the magnetic field during gas compression.
  }
  {
  In this work, we investigate the role of the global magnetic field morphology on a galaxy's $\rm {B-\rho}$ relation, as well as the evolution of the relation over time.
  }
  {
  We develop MHD simulations of Milky-Way-like galaxies including gravity, star formation, and supernova feedback.
  The models take into account non-equilibrium chemistry up to $\rm H_2$ formation, which is used to fuel star formation. We consider two different initial magnetic field morphologies: one completely ordered (toroidal) and the other completely random. In these models, we study the dynamical importance of the magnetic field through the plasma $\beta$ and the $\rm {B-\rho}$ relation.
  }
  {
  For both magnetic morphologies, low-density regions are thermally supported, while high-density regions are magnetically dominated. Equipartition is reached earlier and at lower densities in the toroidal model.
  However, the $\rm B-\rho$ relation is not unique even within the same galaxy, as it consistently includes two different branches for a given density, with $\kappa$ ranging from about 0.2 to 0.8. 
  The mean value of $\kappa$ for each model also displays significant variations over time, which supersede the differences between the two models.
  }
  {
  While our findings suggest that the magnetic field morphology does influence the galactic $\rm B-\rho$ relation, its impact is transient in nature, since time-averaged differences between the models fall within the large temporal scatter.
  The context and time-dependent nature of the $\rm B-\rho$ relation underscore the need for comprehensive research and observations to understand the intricate role of magnetic fields in star formation processes across diverse galactic environments. 
  }
  \keywords{MHD - galaxies: magnetic fields, evolution, ISM - ISM: magnetic fields - stars: formation - methods: numerical}

\maketitle
%

\section{Introduction}
\label{sec:introduction}

Magnetic fields are fundamental for several crucial processes in the interstellar medium (ISM). This influence naturally translates into measurable effects in galaxy evolution \citep[e.g.,][]{pakmor13,voort21,MA2020,Whittingham2021}.
On galactic scales, magnetic fields have the ability to guide the propagation of cosmic rays \citep{fermi,kulsrud,cesarsky,desiati,shukurov} and determine the locations where molecular clouds form, through the interplay between gravity and magnetic buoyancy \citep{Parker1966, Mouschovias1974}.
Very importantly, magnetic fields can significantly influence star formation \citep[see, e.g.][for recent reviews]{hennebelle,pattle}. When the mass-to-flux ratio  ($M_g/\Phi_B$) exceeds a critical threshold, the magnetic support against gravity can slow down or even stop the collapse of a cloud \citep[e.g.,][]{mestel_spitzer,mouschovias_spitzer}. 

To quantify the importance of magnetic fields with respect to other forces several metrics are used, such as the plasma beta ($\beta$) to assess the significance of magnetic fields relative to the thermal pressure. Plasma $\beta$ represents the ratio of thermal pressure, $\rm P_{th}$, to magnetic pressure, $\rm P_{mag}$, and if it is greater than 1 the gas is thermally dominated  \citep[e.g.,][]{pattle}. It is essential to note that accurately measuring the magnetic field strength and the thermal pressure in the ISM can be challenging. 

Similarly, when considering the importance of magnetic fields in relation to gravity, measuring the $\rm M_g/\Phi_B$ in astrophysical systems poses challenges because the complexity of astrophysical environments presents significant obstacles in accurately determining the magnetic flux \citep[e.g.,][]{pattle}.

To overcome the challenges in directly measuring the $M_g/\Phi_B$ and to quantify the importance of magnetic fields, astronomers often resort to the $\rm B-\rho$ relation, which is typically described by a power law. This relation provides insights into how the magnetic field reacts to condensations and offers information about the collapse geometry of molecular clouds \citep{mestel,mouschovias_spitzer,crut10,tritsis15,pattle}. 

\cite{mestel} was the first to derive a theoretical $\rm B-\rho$ relation for a spherically symmetric, isotropically collapsing cloud, based on the mass and magnetic flux conservation:
\begin{eqnarray}
   M_g\propto \rho R^3&=&const \label{eq:mass_cons},\\
   \Phi_B=BR^2&=&const \label{eq:flux_cons},
\end{eqnarray}
where $M_g$ is the gas mass, R the cloud radius, $\rm \Phi_B$ the magnetic flux and $\rm B$ the magnetic field, assumed uniform. With $\rm R \propto\rho^{-1/3}$ from Eq. \ref{eq:mass_cons}, Eq. \ref{eq:flux_cons} gives $\rm B\propto \rho^{2/3}$.

Some years later, \cite{Mousc1,Mouschovias2} derived a $\rm B-\rho$ relation for isothermal self-gravitating clouds. He assumed that the ratio of magnetic to thermal pressure in the deep interior of clouds tends to remain constant and close to unity, $\rm \beta=1$:
\begin{equation}
    \frac{P_{b}}{P_{th}}=\frac{B^2}{8\pi\rho c_s^2}\simeq 1 ,
    \label{eq:pres_equil}
\end{equation}
where $c_s$ the isothermal sound speed. Clearly, from Eq. \ref{eq:pres_equil}, $\rm B\propto\rho^{1/2}$.
Summarizing, the two slopes, 2/3 \citep{mestel} and 1/2 \citep{Mousc1,Mouschovias2}, imply different kinds of compressions for the clouds: the former suggests an isotropic spherical geometry, while the latter implies a slab-like or filamentary geometry; in the 1/2 case, the magnetic field lines are either perpendicular to the slab or inclined relative to the primary axis of the filament.
More generally, if the equation of state is $\rm P\propto\rho^{\gamma}$, then $\rm B \propto \rho^{\gamma/2}$ \citep{mous91}. In the following, we will be referring to the slope $\kappa$ of the $\rm B-\rho$ relation by assuming $\rm B \propto \rho^{\kappa}$.

Since Zeeman measurements of the magnetic field in the atomic and molecular ISM became accessible, the $\rm B-\rho$ relation has also been thoroughly studied in observations. For example, \cite{verschuur} reports that $\kappa =2/3$ based on Zeeman measurements of nine HI clouds, however, without giving an explicit fit. 
Later, \citet{crutcher99} supported $\rm \kappa \simeq 1/2$ by fitting a larger number of cloud data with number densities greater than 100 $cm^{-3}$, and finding $\kappa=0.47 \pm 0.08$.
In a subsequent analysis with a larger number of atomic and molecular regions, \cite{crut10} displayed a log(B)-log(n) diagram that is flat at low densities, indicating no correlation between them, while for higher densities the data follow a trend close to $\rm B \propto \rho^{2/3}$.
However, \cite{tritsis15} revisited the \cite{crut10} data by examining different cloud morphologies; they found that the preferred slope is $\rm \kappa=0.5$ when the entire density range is considered. \cite{li15} fitted an even flatter slope, $\rm \kappa=0.41$ to observations of the massive star-forming region NGC 6334. By extending the Bayesian analysis from \cite{crut10} and by using recent observational and theoretical developments, \cite{jiang20} showed that $\kappa$ cannot be reliably estimated when the observational uncertainty exceeds 2 due to the errors-in-variables bias. Further, \cite{myers21} examined 17 dense cores using the Davis-Chandrasekhar-Fermi (DCF) method to estimate plane-of-sky field strengths $\rm B_{pos}$; they obtained a best-fit value of $\kappa=0.66$ for their dataset. Additionally, \cite{liu} compiled DCF data and obtained a best-fit value of $\rm \kappa=0.57$, but they noted that the relation contains substantial uncertainties.

Despite the great progress of magnetic field observations in the ISM, we are far from fully understanding the connection between the magnetic field and galaxy evolution, mainly because it is inherently difficult to measure the magnetic field strength and to characterize its topology.
In parallel with observational efforts, theorists have been trying to gain insights from numerical simulations. These simulations can provide the $\rm B-\rho$ relation and allow for a comparison with observational data.

Early on, \cite{fiedler92,fiedler93} performed numerical MHD simulations of isothermal, axially symmetric, gravitational contracting cores, predicting a slope of $\rm \kappa=0.47$. \cite{kudoh} developed the first fully 3D simulation to study molecular cloud fragmentation; they found that for higher densities, $\rm \kappa$ tends to be 0.5. \cite{collins} simulated supersonic, super-Alfvénic, and self-gravitating turbulence: throughout the collapsing gas, $\rm \kappa=0.5$ was found. \cite{mocz} simulated supersonic, turbulent, isothermal, self-gravitating gas with a range of magnetic mean-field strengths; they showed that when the kinetic energy dominates over the magnetic pressure, $\rm \kappa=2/3$ implying an isotropic collapse while for a dominant large-scale magnetic field the collapse is anisotropic with $\rm \kappa=0.5$.

In more recent studies, \cite{Seta} investigated the turbulent dynamo in a two-phase medium, obtaining 
$\rm \kappa=0.22$ and $\rm \kappa=0.27$ for a solenoidal turbulent driving at low and high temperatures, respectively; for the compressive case, the slope was found to be $\rm \kappa=0.71$ for low temperatures and $\rm \kappa =0.51$ for higher temperatures, in all cases with a large scatter around the relation. 
In simulations of an isothermal turbulent medium collapse, \cite{brandenburg_ntormousi} found that the scatter around the $\rm B-\rho$ relation was high enough to accommodate various exponents. 
In simulations of decaying turbulence,
\cite{auddy} simulated turbulent molecular clouds and they observed a distinct break density that separates a relatively flat, low-density regime from a power-law regime at higher densities. 
They reported that the transition density increases with increasing values of the Alfvén Mach number.

On galactic scales, numerical models so far reach contradicting results regarding the $\rm B-\rho$ relation.
\citet{wang} modeled the formation and early evolution of disk galaxies with a uniform magnetic field and discovered a flat relation for low densities and a slope of 1/2 for higher densities. However, \citet{pardi}, focusing on kpc-scaled regions within a galactic disk, did not observe an increase in magnetic field with density. Meanwhile, \citet{Girichidis}, adopting a similar setup, reported a slope of 2/3 for low-density gas and 1/4 for high-density gas. Furthermore, conducting cosmological simulations of Milky Way-like galaxies with an initial uniform magnetic field, \citet{ponnada} found results consistent with both slopes of 2/3 and 1/2.

Until now, studies investigating the $\rm B-\rho$ relation have primarily assumed a uniform magnetic field, with little exploration of different magnetic field topologies. Existing research focusing on kpc- and pc-sized regions consistently indicates that increasing magnetic field strength suppresses the dense gas fraction and overall star formation rate within the model \citep{myers, federrath, pardi, Girichidis, wuster}. However, recent simulations have revealed that a stronger random magnetic field can actually
lead to a faster collapse in dense regions \citep{brandenburg_ntormousi}.

In this paper, we investigate the impact of a galaxy's initial magnetic field topology on the $\rm B-\rho$ relation over time, specifically exploring the possibility that different field morphologies induce different compression modes ($\kappa$ values). To achieve this, we perform two galaxy simulations, each initialized with a different initial magnetic field: one with an ordered toroidal field and the other with a random magnetic field.

We describe the numerical code and the setup in Sect.~\ref{sec:methods}. The results of our investigations are presented in Sect.~\ref{sec:results}, a discussion in Sect.~\ref{sec:discuss}, and the conclusions in Sect.~\ref{sec:conclus}.

\section{Method and setup}\label{sec:methods}

We perform MHD simulations of two Milky Way (MW)-sized disk galaxies, accounting for self-gravity, star formation, supernova feedback, and chemistry. 
The initial conditions comprise a dark matter (DM) halo, a gaseous halo, a gas disk, and a stellar disk. 
The simulations achieve a spatial resolution of 24 pc in high-density regions, providing detailed information at that scale. 

\subsection{Magneto-hydrodynamics}\label{ssec:code}

The galaxy simulations were conducted with RAMSES, an Eulerian, magneto-hydrodynamics (MHD), Adaptive Mesh Refinement (AMR) code (\citealt{teyssier2002}; \citealt{fromang}). The models include dark matter and stars, represented by collisionless particles, and a multiphase gaseous component, treated as a magnetized fluid. 
RAMSES solves the ideal MHD equations in the following form: 
\begin{equation}
    \frac{\partial \rho}{\partial t}+\nabla(\rho \textbf{v})=0 ,
\end{equation}
\begin{equation}
    \frac{\partial (\rho \textbf{v})}{\partial t}+\nabla(\rho \textbf{v} \textbf{v} - \textbf{B} \textbf{B})+ \nabla P_{tot}=-\rho \nabla \phi , 
\end{equation}
\begin{equation}\label{eq:energy_evolution}
    \frac{\partial E_{tot}}{\partial t}+\nabla([(E_{tot}+P_{tot})\textbf{v}-(\textbf{v} \cdot \textbf{B}) \cdot \textbf{B})]=-\textbf{v} \cdot \nabla \phi -\rho \Lambda + \Gamma ,
\end{equation}
\begin{equation}
        \frac{\partial \textbf{B}}{\partial t}-\nabla \times (\textbf{v} \times \textbf{B})=0 ,
\end{equation}
where $\textbf{v}$ is the velocity, $\phi$ the gravitational potential, $\Lambda$, and $\Gamma$ the cooling and heating rates of the gas respectively.
RAMSES solves the MHD equations using a Godunov scheme with Constrained Transport to guarantee that the solenoidality condition for the magnetic field {\bf ($\nabla \cdot \textbf{B} =0$)} is fulfilled.

\subsection{Non-equilibrium chemistry}\label{ssec:krome}

The complex chemistry of interstellar gas plays a vital role in its thermal evolution. Properly modeling the non-equilibrium chemistry of molecular hydrogen formation and dissociation also leads to a more accurate treatment of star formation \citep{valdivia,decataldo}. 
For this reason, we use a customized version of RAMSES that follows the non-equilibrium chemistry of H$_2$ formation and dissociation through the KROME package \citep{grassi}.
The implemented chemical network contains H, H$^+$, H$^-$, H$_2$, H$_2^+$, He, He$^+$, He$^{++}$, and electrons and allows us to track the evolution and thermodynamics of ionized, atomic and molecular gas \citep{bovino}, and it has been adopted both in galaxy \citep{pallo} and molecular clouds simulations \citep{decataldo}.
Various heating and cooling processes regulate the thermal state of the gas (eq. \ref{eq:energy_evolution}). In addition to typical atomic processes, our simulations also incorporate the cooling resulting from the presence of $\rm{H_2}$ in the gas phase, and metal cooling.
Here we have assumed solar metallicity for the ISM gas.

The radiation field can modify the chemical evolution by causing the ionization and dissociation of atoms and molecules. 
On-the-fly radiative transfer is not explicitly included in our simulations. Instead, we allow for a uniform radiation background of 1 $G_0$ where $\rm{G_0 = 1.6 \times 10^{-3} \; erg \,cm^{-2} \,s^{-1}}$ is the far Ultraviolet (UV) flux in the Habing band (6 - 13.6 eV) normalized to the average MW value \citep{habing}. The spectral shape is that of a non-ionizing \citet{draine} distribution, which is appropriate to describe a MW-like interstellar radiation field, and we select ten photon bins to cover the energy range from 5 to 13.6 eV,
Note that self-shielding of molecular hydrogen from the photo-dissociating Lyman-Werner radiation is included adopting a cell-by-cell prescription following \citet{richings:2014}.

\subsection{Star formation and supernova feedback}\label{ssec:stars}

Our model includes star formation and supernova feedback. The star formation recipe is based on the Schmidt-Kennicut relation \citep{schmidt,kennicutt} following the $\rm H_2$ density \citep{pallo}.
During star formation, a fraction of the molecular mass of a cell transforms into a new stellar particle with a stellar mass $\rm M_{*,c}$. This process occurs with an efficiency $\epsilon_{SF}$ that we set to $\rm\epsilon_{SF}=1\%$,  which is reasonable for MW-like galaxies, and we kept it constant between the two models to facilitate their comparison. The random sampling from a Poisson distribution was originally implemented by \cite{rasera}. 
To prevent the spurious formation of stellar particles, a minimum stellar mass of $\rm M_{*,c}=10^3\msun$ is set, which represents the smallest stellar mass that can form in a cell.

The supernova feedback is simulated by injecting thermal energy into the 27 neighboring cells around the star particle. 
These 27 cells consist of the central cell containing the supernova, which we can call $C_i$, the 6 cells that share a face with $C_i$, 12 cells that share an edge with $C_i$, and 8 cells that share a vertex with $C_i$. This arrangement ensures that the explosion is represented in a numerically stable way across varying resolution levels.
Each supernova produces $10^{51}$ erg. In our simulations, 20$\%$ of the stellar population in a particle undergoes supernova explosions 3 Myr after the star particle's generation. The initial stellar disk, which is composed of older stars, does not contribute to the supernova budget. Note that radiation or wind feedback from the stellar particles is not included in our simulations.

\subsection{Initial conditions}\label{ssec:IC}

We generate the initial conditions for the dark matter, stars, and gas using the DICE code \citep{perret}. DICE can combine the different components of a galaxy and sample three-dimensional (3D) distributions of particles using a Markov chain Monte Carlo (MCMC) algorithm. The dynamical equilibrium is reached by solving the Jeans equations.

We prepare MW-like galaxies at redshift $z=0$ with a virial velocity of 200 km/s, which corresponds to a total mass of $M_{\rm tot} = 10^{12} M_{\odot}$. The galaxy is composed of a DM halo (97.5$\%$ of $M_{\rm tot}$), a thin stellar disk (1.425$\%$), a gaseous disk (0.075$\%$), and a gaseous halo (1$\%$). 
The DM and gaseous haloes follow a pseudo-isothermal density profile with a scale length of 3 kpc and a radial density cut at 100 kpc. Each DM particle has a mass of $4.4\times 10^5 \; M_{\odot}$. The thin stellar disk follows a Miyamoto-Nagai profile \citep{miyamoto} with a scale length of 3 kpc and a radial density cut at 12 kpc while all the initial stellar particles have the same mass. The gaseous disk follows an exponential density model with a scale length of 4 kpc, a radial density cut at 15 kpc and a constant density vertical profile cut at 0.75 kpc. The initial temperature is 8000 K without an initial turbulent velocity field. 

For these simulations we use an AMR with a coarse resolution of $128^3$ and five levels of refinement, by adopting the following strategy. We begin with a geometry-based refinement, 
where the two first levels are triggered in two cylindrical regions of decreasing size that are centered on the galaxy and fully encompass the gaseous disk with densities $\rm n/cm^{-3}>10^{-3}$ at all times. 
Inside the inner region, three additional levels of refinement are triggered by a Jeans-based criterion, i.e. we require the local Jeans length to be resolved with at least 10 cells. This strategy results in
the highest effective resolution of $4096^3$ in the densest
regions. The simulation box is a cubic volume with a side length of 100 kpc, so the maximum resolution corresponds to 24 pc.

It is important to note that star formation is enabled at around 100 Myr for both models, to allow the gas to collapse and form dense clumps in the disk. The simulations run for a total duration of 500 Myr.

\begin{figure*}
    \centering
    \includegraphics[trim={0cm 0cm 0cm 0cm},clip,width=.44\linewidth]{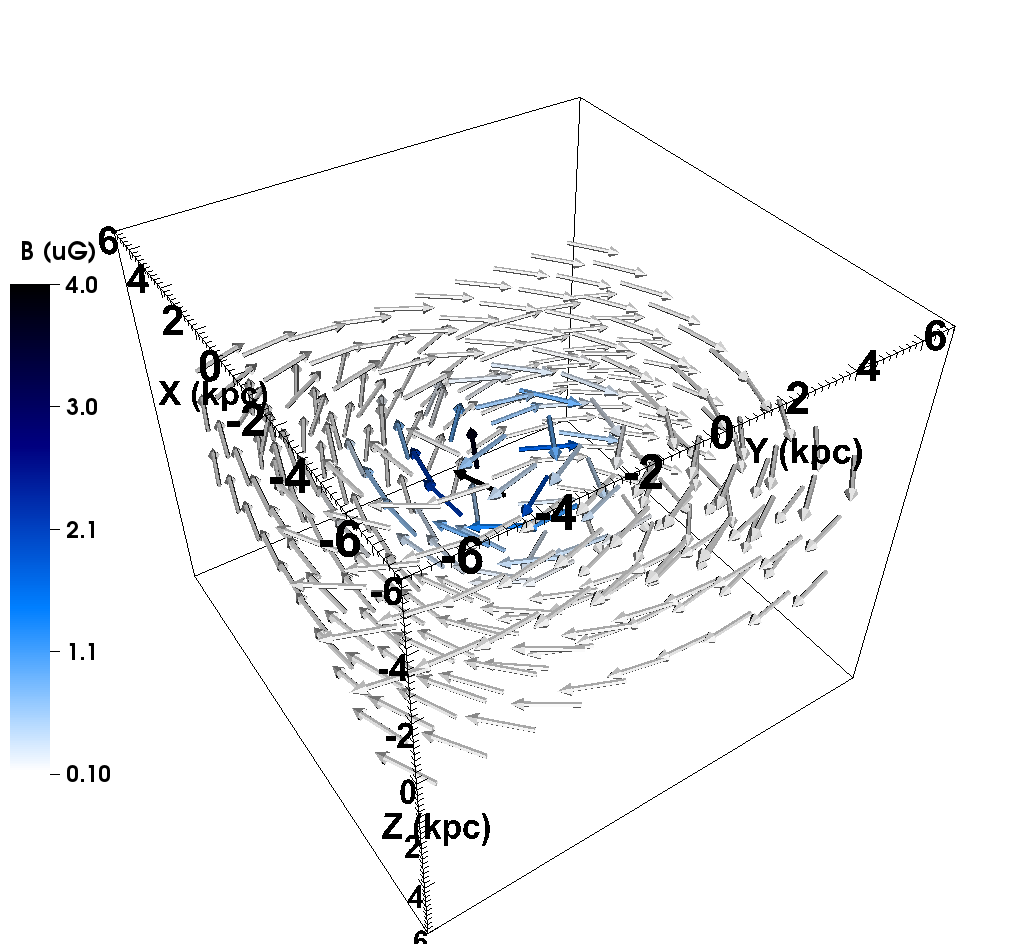}
    \includegraphics[trim={0cm 0cm 0cm 0cm},clip,width=.46\linewidth]{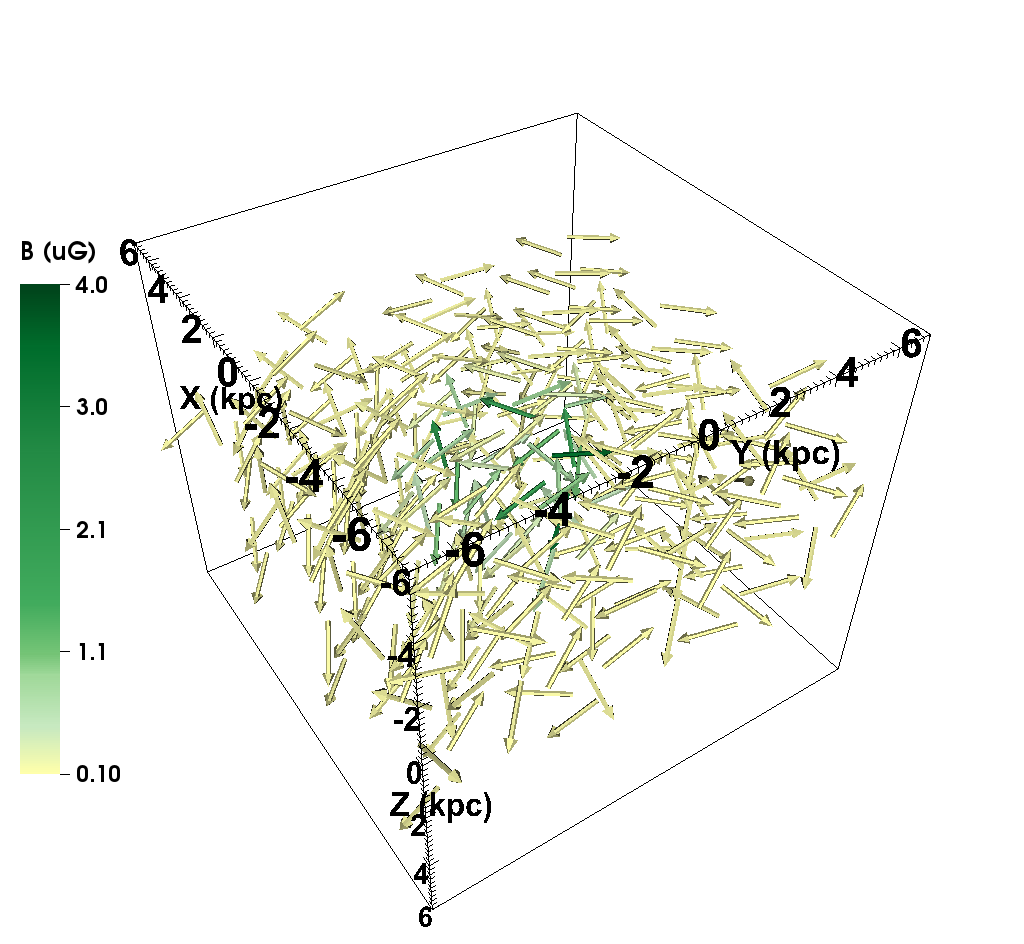}
    \caption{3D representation of the initial conditions ($t=0$) for the magnetic field vectors for the two simulations, showing the toroidal (T) and random (R) cases on the left and right side, respectively. For illustration purposes, we show only vectors in the central region of the simulation.}
    \label{fig:init}
\end{figure*}

The initial conditions of the two galaxies are identical, except for the magnetic field morphology. One galaxy, referred to as model T, is initialized with a toroidal magnetic field, while the other galaxy, referred to as model R, starts with a random magnetic field. In both models, the magnetic field has an exponentially declining profile.
The initial magnetic field topologies are shown in a 3D representation of magnetic field vectors in Fig.~\ref{fig:init} for both models.

The initial magnetic field is generated by specifying the vector potential $\rm{\textbf{A}}$, and then calculating the magnetic field by taking the curl of $\rm{\textbf{A}}$, $\textbf{B} = \nabla\times{\textbf{A}}$. This method ensures that $\rm{\nabla\cdot\textbf{B}=0}$ to machine precision.
For model T the definition of $\rm{\textbf{A}}$ follows:
\begin{equation}
   \textbf{A}\propto \hat{e}_z \exp\left(-r/R_0\right)~\exp\left(-z/H_0\right) \,,
\end{equation}
where $\hat{e}_z$ is the unit vector along the $z$ axis, $r$ represents the cylindrical radius, and $z$ is the vertical height from the galactic plane. The parameters $R_0$ and $H_0$ represent the scale length and scale height of the magnetic field, respectively. In this work, the values of $R_0$ and $H_0$ are set to 1 kpc, and the magnetic field strength is normalized such that it is 10 $\mu$G at the galactic center.

For model R, we create a vector potential in Fourier space using a power spectrum:
\begin{equation}
    P(\tilde{A_i})\propto k^{-4} ,
\end{equation}
where $\rm i=x,y,z$ represents the component of $\rm{\textbf{A}}$. Each component of $\rm{\textbf{A}}$ is then obtained in physical space through an inverse Fast Fourier Transform. The resulting magnetic field is normalized to match the central strength of model T and is also convolved with the same exponential profile of radius and height as model T.

\section{Results}\label{sec:results}

We study the evolution of the morphology of the galaxies, their gas and stellar content, and their star formation histories (Sec. \ref{ssec:morphology_sfr}).
Then, we examine the thermodynamical and magnetic state of the gas, in particular investigating the plasma beta relation in different gas phases (Sec. \ref{ssec:beta}).
Finally, we focus on the time evolution of the \rm B-$\rho$ relations (Sec. \ref{ssec:B_rho}).

\subsection{Morphology and star formation rates of galaxies}\label{ssec:morphology_sfr}

\begin{figure*}
    \centering
    \hspace{-2.2cm} Model T  \hspace{5.5cm} Model R \\
    \rotatebox{90}{\hspace{2.7cm} t = 200 Myr}
    \includegraphics[trim={0cm 0cm 6.5cm 0cm},clip,width=.4\linewidth]{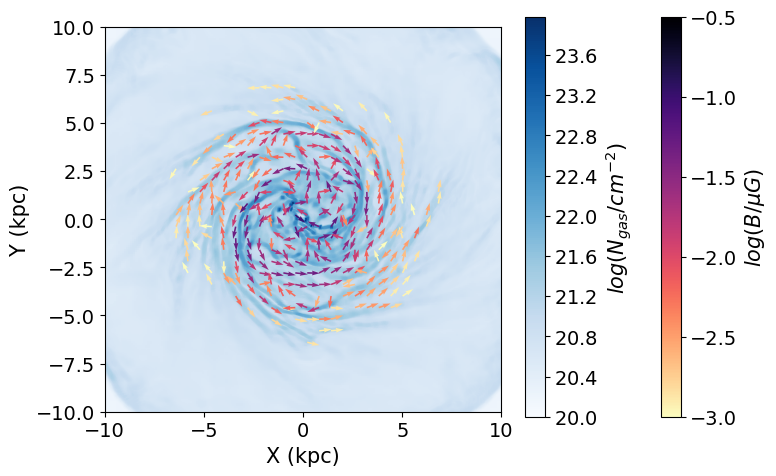} 
    \includegraphics[trim={1cm 0cm 0cm 0cm},clip,width=.57\linewidth]{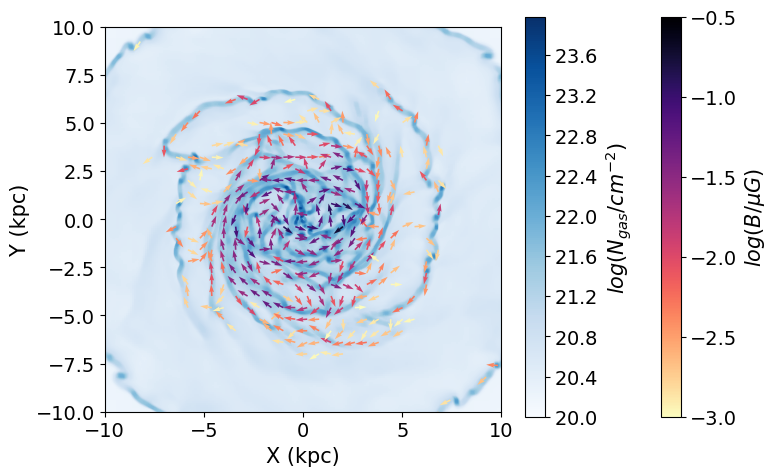}\\
    \rotatebox{90}{\hspace{2.7cm} t = 300 Myr}
    \includegraphics[trim={0cm 0.cm 6.5cm 0cm},clip,width=.4\linewidth]{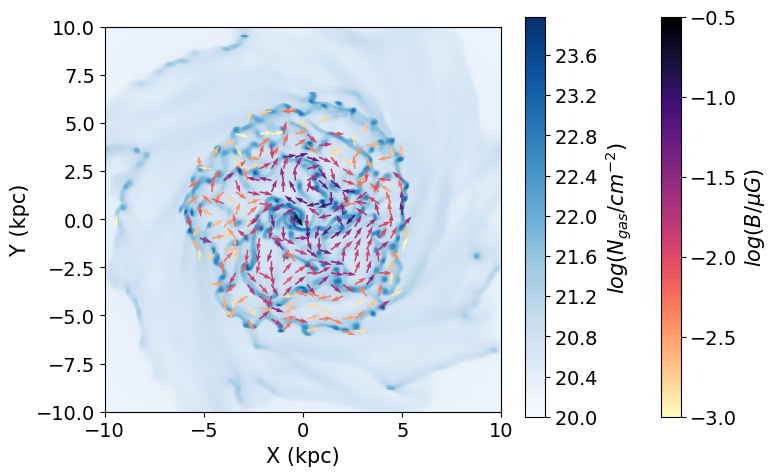} 
    \includegraphics[trim={1cm 0cm 0cm 0cm},clip,width=.57\linewidth]{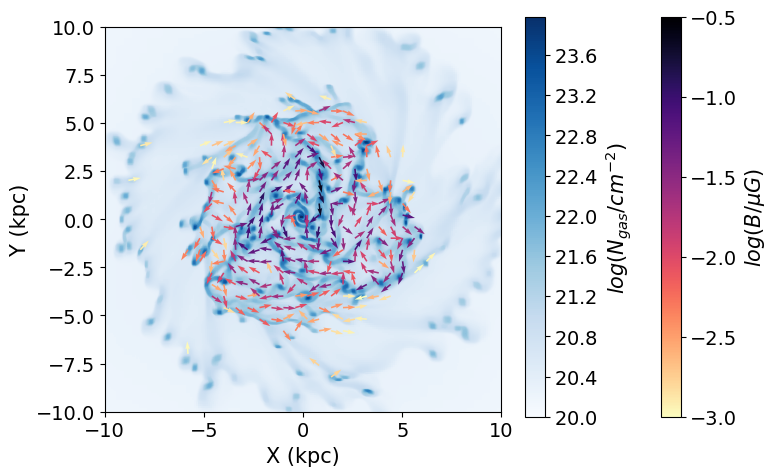}\\
    \rotatebox{90}{\hspace{2.7cm} t = 500 Myr}
    \includegraphics[trim={0cm 0cm 6.5cm 0cm},clip,width=.4\linewidth]{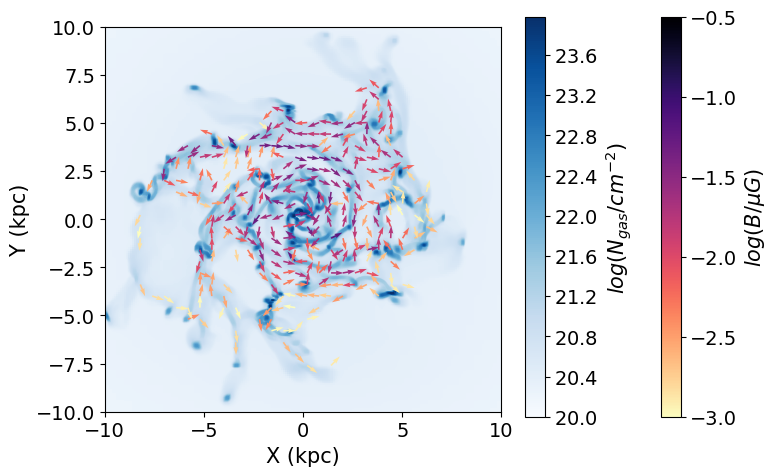} 
    \includegraphics[trim={1cm 0cm 0cm 0cm},clip,width=.57\linewidth]{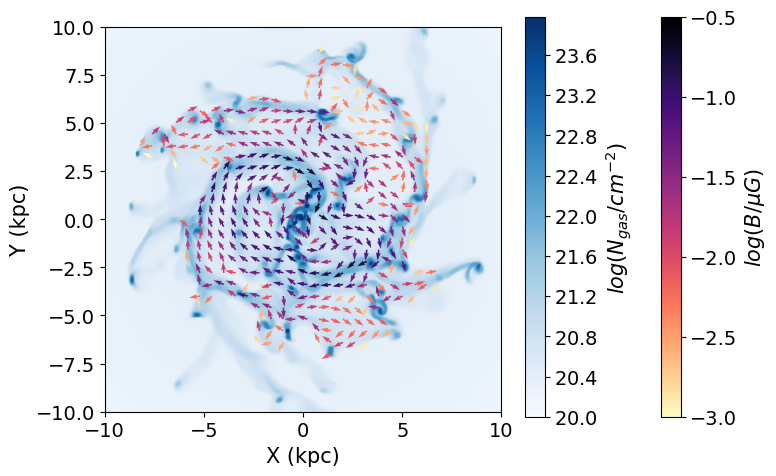}
    \caption{Face on maps of the gas column density ($N_{gas}$) and the projected magnetic field vectors ($B$). Model T and model R are shown on the left-hand side right-hand side, respectively. From top to bottom we plot the maps at t = 200, 300, and 500 Myr.}
    \label{fig:col_den}
\end{figure*}

Fig.~\ref{fig:col_den} shows face-on total gas column density maps and projected magnetic field vectors for the two galaxies at different times ($t = 200,\, 300,\, \rm and\, 500\, Myr$). Model T is shown on the left-hand side, and model R is on the right-hand side. The magnetic field vectors are color-coded according to their strength. 

At early times, both models exhibit stronger magnetic fields in their central regions, as set in the initial conditions. However, as time progresses, we observe distinct differences in the magnetic field evolution between the two models.
Specifically, model R displays a wider region of strong magnetic field, spanning approximately 0-5 kpc. In contrast, model T shows a narrower range, covering approximately 0-3 kpc, as visually inspected.

\begin{figure*}
    \centering
    \includegraphics[trim={0cm 0cm 0.1cm 0.3cm},clip,width=.47\linewidth]{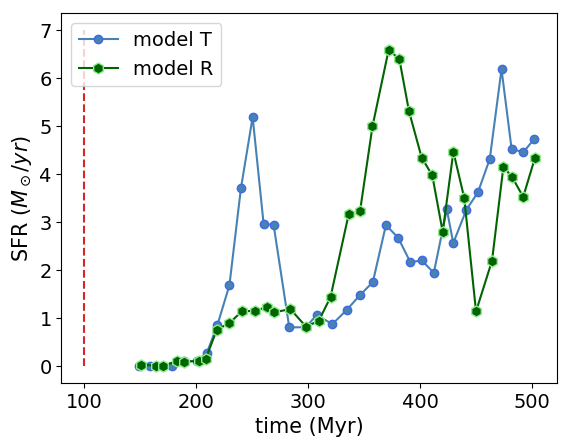} 
    \rotatebox{90}{\hspace{2.8cm} log($M_\star^{\rm new}/M_{\odot})$}
    \includegraphics[trim={0.8cm 0cm 0cm 0cm},clip,width=.45\linewidth]{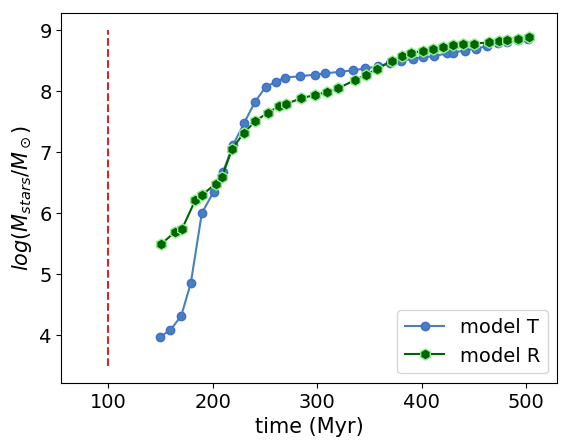}
    \caption{Star formation rate (SFR, left-hand side) and cumulative mass of the new stars ($M_\star^{new}$ right-hand side) as a function of time of model T (blue) and model R (green). The red dashed line represents the time when star formation is allowed in the simulation.}
    \label{fig:m_t}
\end{figure*}

As we outlined in the introduction, star formation is sensitive to the strength and morphology of the magnetic field. Therefore, the sharp difference in magnetic energy distribution between the two models that we notice on visual inspection could in principle affect the star-forming properties of the galaxies.
In Fig.~\ref{fig:m_t} we plot the star formation rate (SFR, left-hand side) and the cumulative mass of the new stars (right-hand side) as a function of time. The SFR history is generated with a time binning of 10 Myr.
We have checked that varying this time interval does not affect the resulting trend of SFR with time.
The red dashed line indicates the moment where we switch on star formation in the simulation.

We observe some differences between the two galaxies. In model T, the peak of the star formation rate occurs at around 250 Myr and reaches approximately $5.5\,\msunyr$; instead, in model R, the peak of the star formation rate occurs later, at approximately 380 Myr, with a higher value of around $\rm 6.5\,\msunyr$.
Despite the temporal offsets observed in the SFRs between the two models, both ultimately result in a similar total mass in stars, approximately around $\rm 10^9 \msun$ while the total stellar mass from the initial conditions was $\rm 1.4\times 10^{10}\msun$. Furthermore, both models exhibit a comparable average star formation rate, 2.19 $\pm$ 1.69 $\rm M_{\odot}$/yr for model T and 2.32 $\pm$ 1.95 $\rm M_{\odot}$/yr for model R. 
These SFR values align with observational estimates of the SFRs 
in Milky Way-like galaxies, 
\citep{licquia,fraser,boardman,elia}; it is worth noting that Galactic observations also indicate temporal variations in the SFR \citep{lee16}.

\begin{figure*}
    \centering
    \hspace{-0.2cm} Model T  \hspace{6.5cm} Model R \hspace{3.4cm}\\
    \rotatebox{90}{\hspace{3.cm} t = 250 Myr}
    \includegraphics[trim={0cm 0cm 3.cm 0cm},clip,width=.45\linewidth]{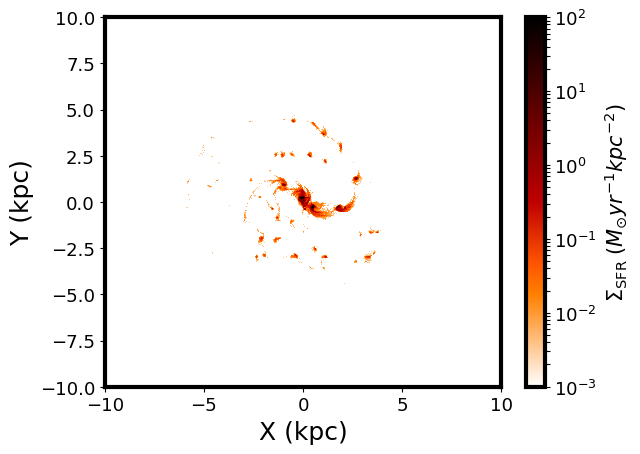} 
    \includegraphics[trim={1.cm 0cm 0cm 0cm},clip,width=.52\linewidth]{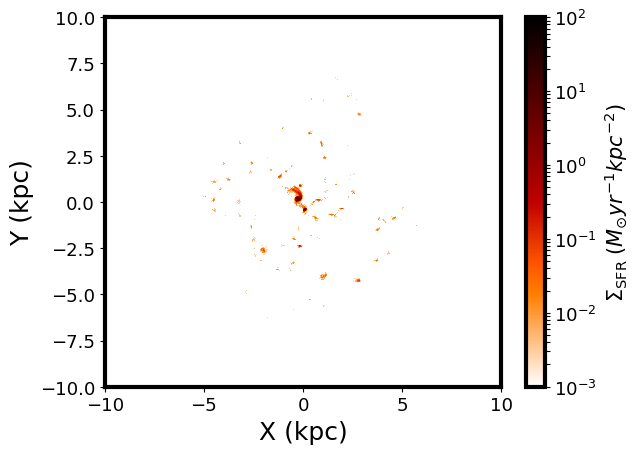}\\
    \rotatebox{90}{\hspace{3.cm} t = 380 Myr}
    \includegraphics[trim={0cm 0cm 3.cm 0cm},clip,width=.45\linewidth]{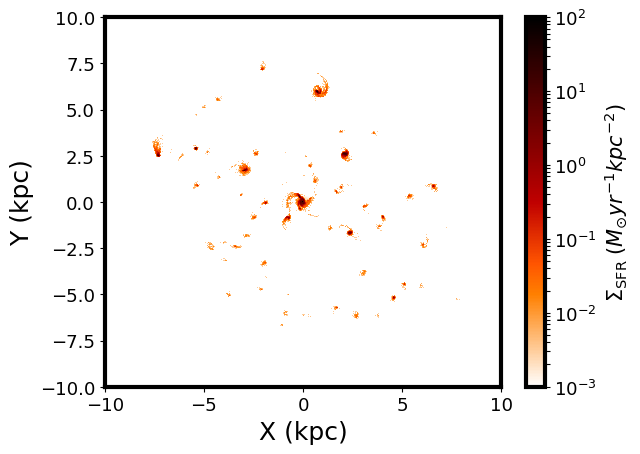} 
    \includegraphics[trim={1.cm 0cm 0cm 0cm},clip,width=.52\linewidth]{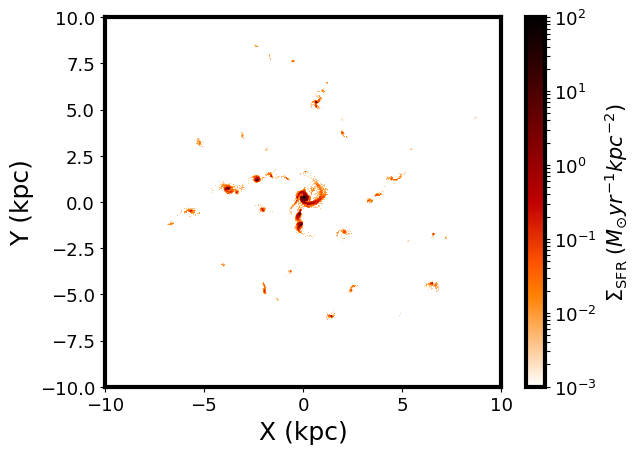} 
    \caption{Maps of the star formation surface density ($\Sigma_{\rm SFR}$) for model T (left-hand side) and model R (right-hand side) at their peak of star formation rate, which happens at t = 250 and 380 Myr, respectively.
    \label{fig:sfr_map_max_min}
    }
\end{figure*}

In Fig.~\ref{fig:sfr_map_max_min}, we show the surface density of the SFR at the time when each model reaches the peak of its SFR, as obtained from Fig.~\ref{fig:m_t} (250 and 380 Myr, for models T and R respectively). Overall, star formation is more intense in the central 0.5-1~kpc of the galaxies and in spiral structures at larger radii reaching peaks of 1.41 and 1.85 $\rm M_{\odot} yr^{-1}kpc^{-2}$ for model T and model R at 250 and 380 Myr, respectively. The corresponding mean surface density of the star formation rate is $8\times 10^{-4}$ and $7.3\times 10^{-4} \rm M_{\odot}yr^{-1}kpc^{-2}$.
At 250 Myr, star formation in both models is mainly organized in clumps, however, model T includes a smooth central region of about 5 kpc of high star formation rate.

\subsection{Thermodynamical and magnetic state of the gas}\label{ssec:beta}

\begin{figure*}
    \centering
    \hspace{-.6cm} Model T  \hspace{5.5cm} Model R \hspace{4.5cm}\\
    \includegraphics[trim={0cm 0.3cm 5.1cm 1cm},clip,width=.45\linewidth]{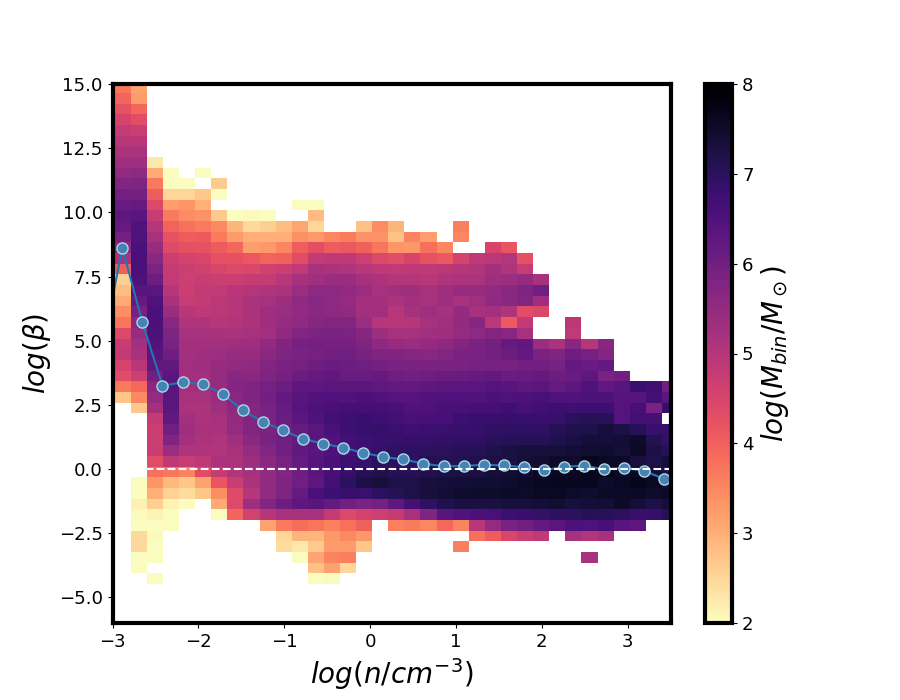} 
    \includegraphics[trim={2.8cm 0.3cm 1.5cm 1cm},clip,width=.47\linewidth]{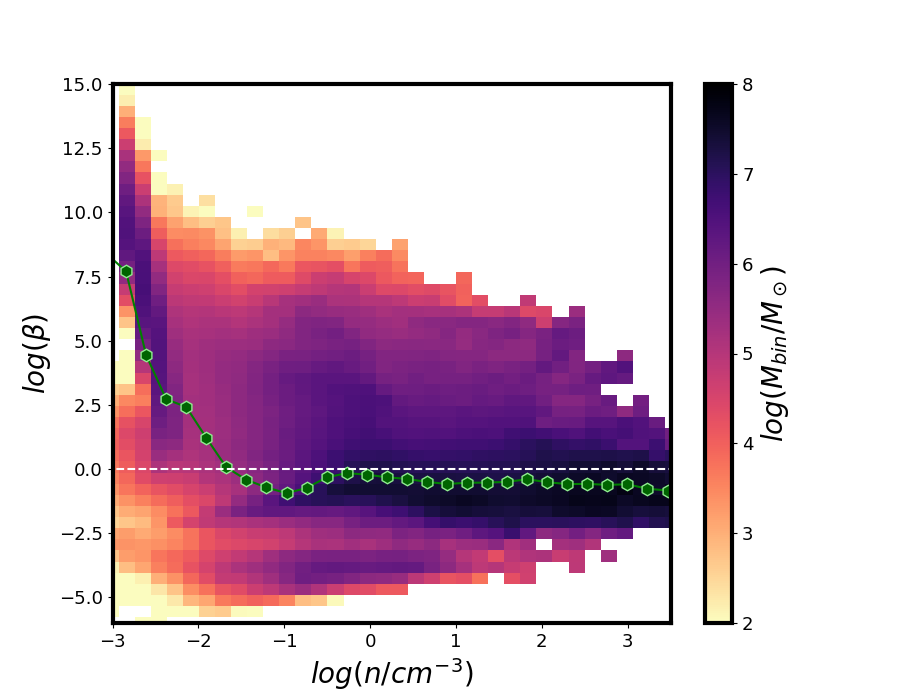}\\
    \includegraphics[trim={0cm 0.3cm 5.1cm 1cm},clip,width=.45\linewidth]{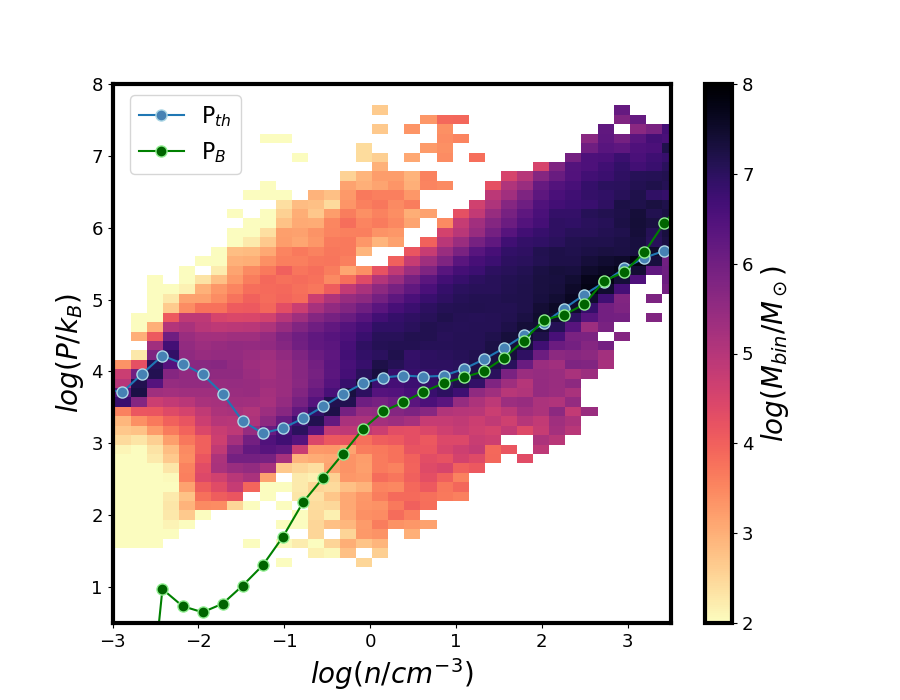} 
    \includegraphics[trim={2.8cm 0.3cm 1.5cm 1cm},clip,width=.47\linewidth]{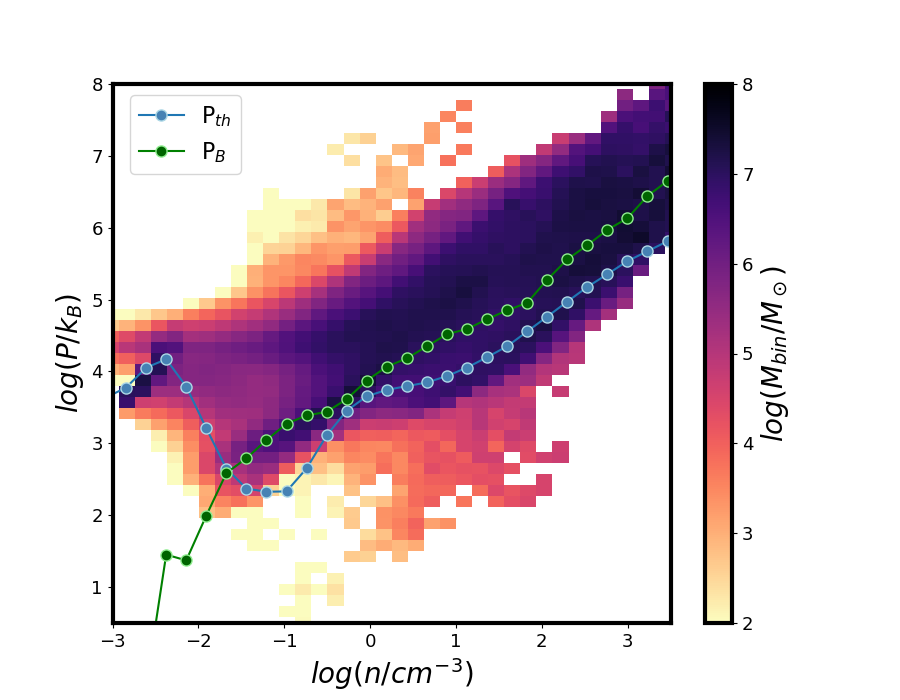}
    \includegraphics[trim={0cm 0.3cm 5.1cm 1cm},clip,width=.45\linewidth]{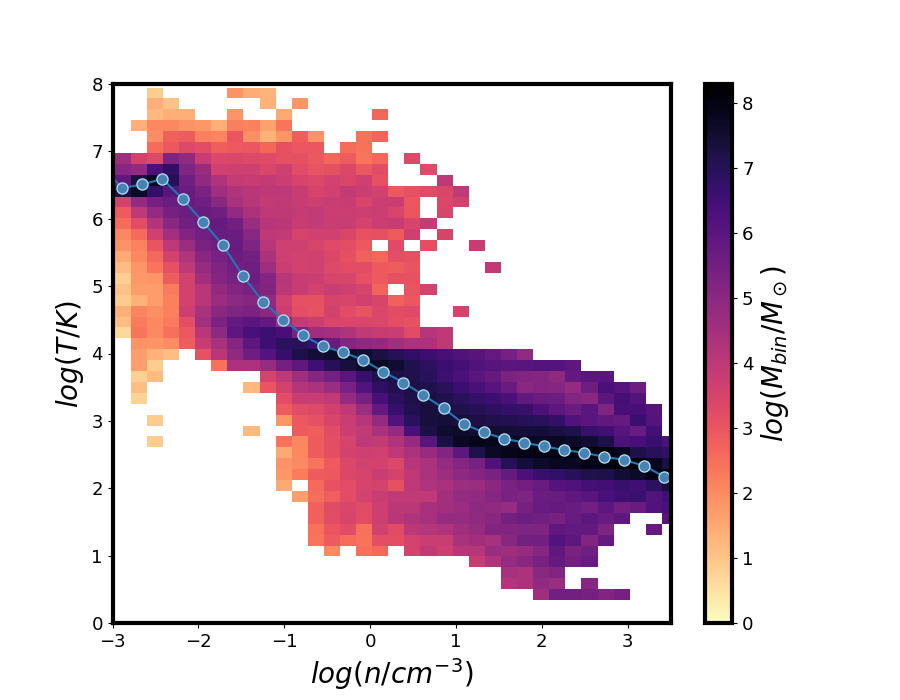} 
    \includegraphics[trim={2.8cm 0.3cm 1.5cm 1cm},clip,width=.47\linewidth]{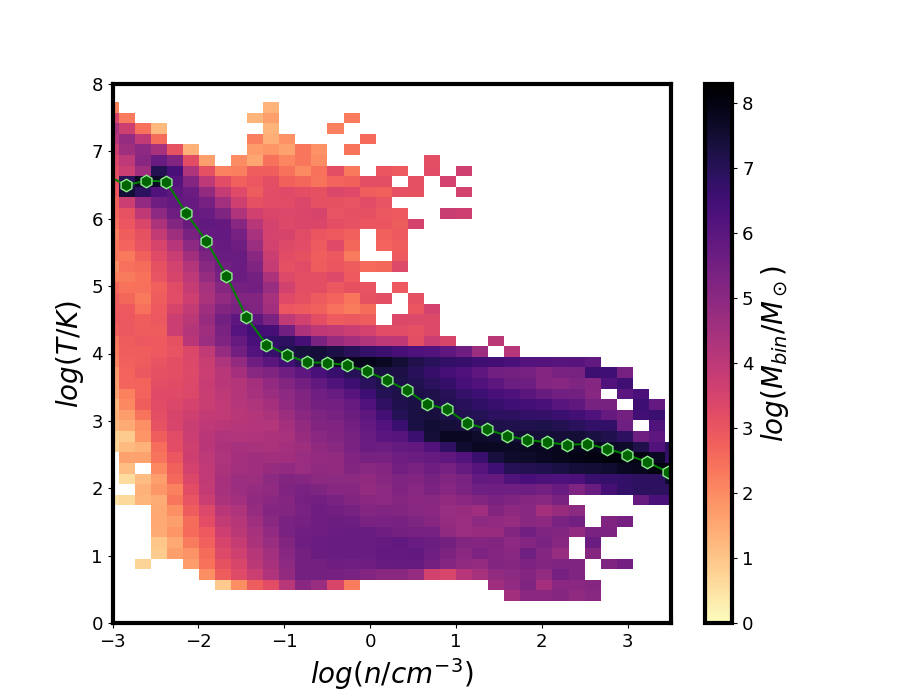}
    \caption{2D mass-weighted histograms of the logarithm of the plasma beta ($\beta$, top), total pressure ($P$, middle), and temperature ($T$, bottom) of the gas plotted against its density (n) for model T (left-hand side) and model R (right-hand side) at t = 500 Myr. The dots represent the mean values per density bin. A white dashed line on the top panel indicates the equipartition of thermal and magnetic pressure.}
    \label{fig:beta}
\end{figure*}

In order to study the dynamical state of the gas in different phases, we first examine the plasma $\rm\beta$ as a function of gas density and time.
The relation between $\rm\beta$ and mean number density n is shown in the top panel of Fig.~\ref{fig:beta} for the final snapshot of each model at 500 Myr. In this figure, the dots represent the mean values of $\rm log ~\beta$ in each density bin. To produce these plots we take into account only the gas cells with $\rm|z|\leq1.5~kpc$ and radius $\rm R\leq13~kpc$ encompassing the entire disk. \footnote{We experimented with different cutoff values at $\log(n/{\rm cm}^{-3})=-1,\, -2,\,{\rm and}\, -3$ to determine the radius, but the results remained consistent. The same was observed with the choice of height, where we also tested considering only the region smaller than the scale height, which is 0.5 kpc.}
We study the behavior of the gas by categorizing it into three distinct phases defined by number density: low-density ($\rm 10^{-2} < n/cm^{-3} < \rm 1$), intermediate-density ($\rm 1 < n/cm^{-3} < \rm 10^{2}$), and high-density phase ($\rm 10^{2} < n/cm^{-3} < \rm 10^{4}$).

In model T, the gas in the low-density phase exhibits high values of $\rm \beta$; at intermediate and high densities, the mean behavior suggests a transition toward equipartition. Furthermore, the regions with high mass bins are predominantly located in the magnetically dominated region ($\rm log(\beta)<0$).

In model R, the mean behavior in the intermediate and high-density phases points to magnetically dominated gas.
Interestingly, there is also a substantial fraction of very magnetically dominated gas ($\rm \beta<-2.5$) in the density range from $10^{-2}$ to $10\,\cc$. Note that the initial conditions consist of only thermally dominated gas with low-density a temperature of $T = 8000\,\rm K$. However, after 100 Myr, the gas is able to cool and condensate at intermediate densities, becoming magnetically dominated.

In order to understand the origin of this behavior, in the middle panel of Fig.~\ref{fig:beta} we plot the total pressure ($\rm log(P/k_B)$). Here, the blue and green dots represent the mean values of the thermal and magnetic pressure per density bin, respectively\footnote{Additional phase diagrams for different times can be found in App.~\ref{appendix:a} (see in particular Fig.~\ref{fig:p_n_t}).}.

In the low-density medium of model T, thermal pressure dominates and the signature of the thermal instability \citep{field1965} is evident from the characteristic bend of the average pressure line \citep{wolfire1995}. The two contributions to pressure are equal in the higher-density regions.
In model R, the magnetic pressure dominates over almost the entire density range, with the exception of very low-density gas ($\rm n<10^{-2}\,\cc$). As a result, the two atomic phases (warm neutral medium (WNM, $\rm -1 < log(n/cm^{-3}) < 0$) and cold neutral medium, (CNM, $\rm 1 < log(n/cm^{-3}) < 2$)) created by the thermal instability are both supported by magnetic pressure.

This magnetic support yields a visible signature in the gas temperature, which is shown\footnote{In App. \ref{appendix:a}, Fig. \ref{fig:T_rho_t} shows similar plots for different times} as a function of density in the bottom panel of Fig.~\ref{fig:beta}. Model R contains a significant gas fraction at low densities ($\rm 10^{-1}<n/\rm cm^{-3}<10$) with temperatures lower than $100\,\rm K$. This phase is absent in the temperature-density plot of model T. The additional magnetic pressure in model R allows low-density gas to cool while still maintaining pressure equilibrium with its surroundings. 

\begin{figure*}
    \centering
    \hspace{0.6cm} Model T  \hspace{6.7cm} Model R \hspace{3.5cm}\\
    \includegraphics[trim={0cm 0cm 0.1cm 0cm},clip,width=.45\linewidth]{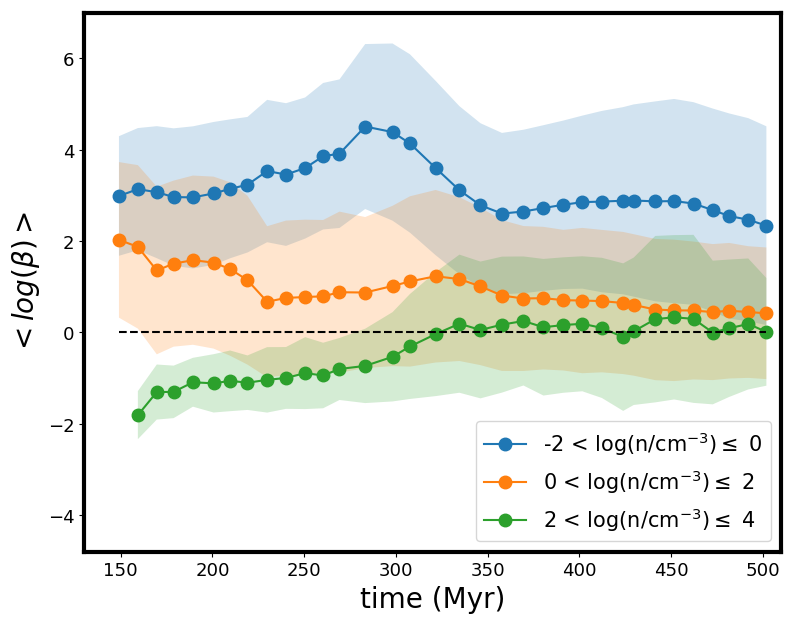} 
    \includegraphics[trim={1cm 0cm 0cm 0cm},clip,width=.43\linewidth]{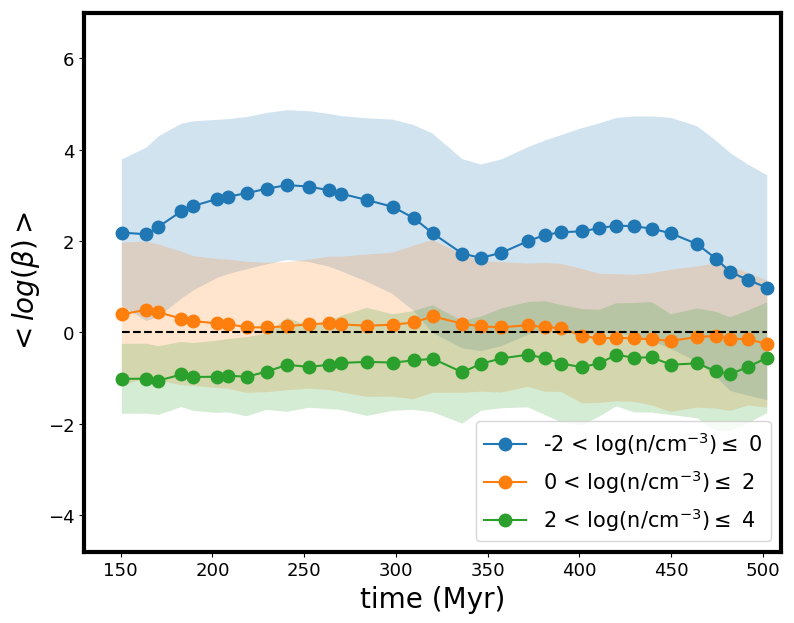}\\
    \includegraphics[trim={0cm 0cm 0.1cm 0cm},clip,width=.45\linewidth]{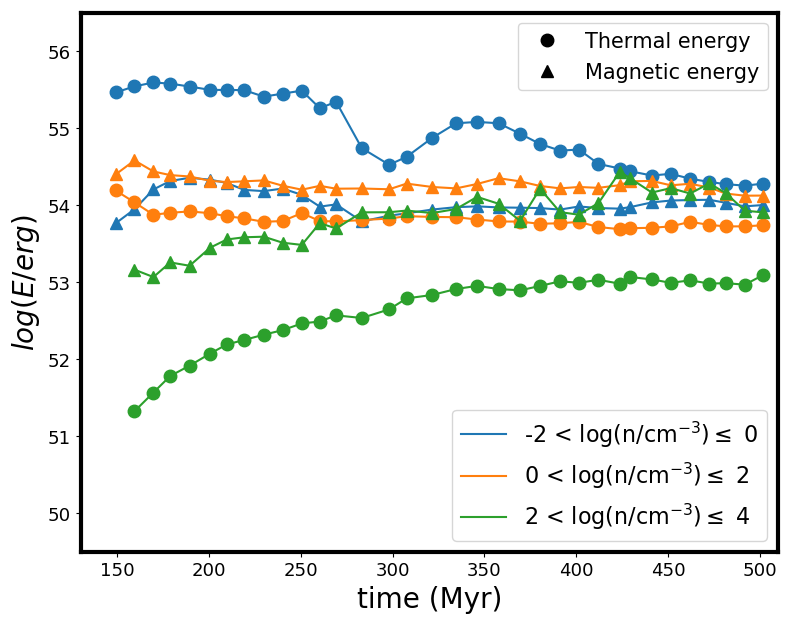} 
    \includegraphics[trim={1cm 0cm 0cm 0cm},clip,width=.43\linewidth]{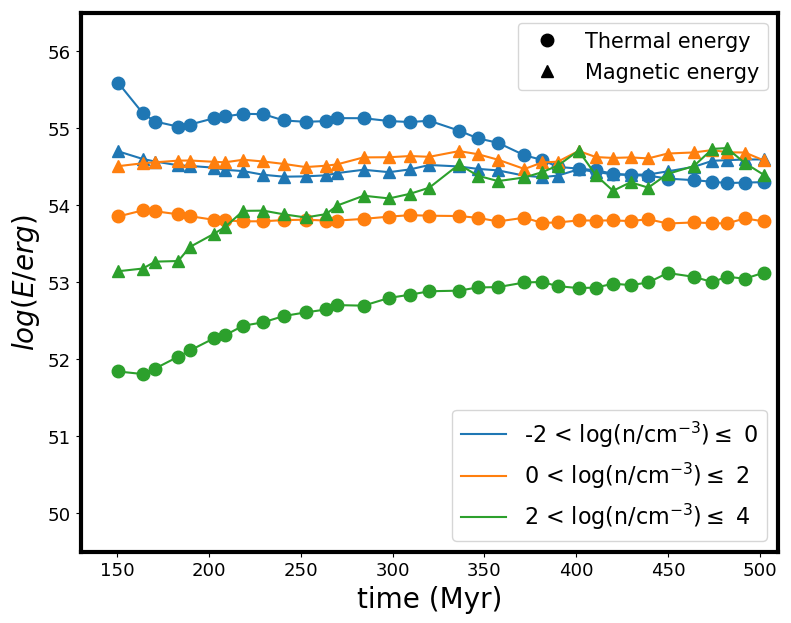}
    \caption{Evolution of $\langle\log\beta\rangle$ (top) and total energies (bottom) for three different phases of the gas. The blue, orange, and green lines show the gas in a low ($-2 < log(n/cm^{-3}) \leq 0$), medium ($0 < log(n/cm^{-3}) \leq 2$) and high ($2 < log(n/cm^{-3}) \leq 4$) density phase, respectively, for model T (left-hand side) and model R (right-hand side).
    In the top plots, the black dashed line corresponds to $\langle\log\beta\rangle=0$, which is the critical value. Above this value, the gas is thermally dominated, while below it, it is magnetically dominated.
    For the bottom plots, circles represent the thermal energy while triangles correspond to the magnetic energy.
    The red dashed line represents the onset of star formation. The shaded regions correspond to the standard deviation.
    \label{fig:beta_t} 
    }
\end{figure*}

In order to study the time evolution of the thermal state of the gas, in Fig.~\ref{fig:beta_t} we plot $\langle\log\beta\rangle$ (top panels) and the total thermal and magnetic energies (bottom panels) as a function of time for the low, intermediate, and high-density phases\footnote{As in Fig.~\ref{fig:beta}, we take into account only the gas with $\rm|z|\leq1.5~kpc$ and radius $\rm R\leq13~kpc$, i.e. within the disk.}.

The blue line represents the low-density gas, the orange line represents the intermediate-density gas, and the green line represents the high-density gas for model T (left-hand side) and model R (right-hand side). In the top plots, the black dashed line corresponds to log($\beta$) = 0, while the red dashed line represents the onset of star formation. 
The reason why we show the calculation of $\beta$ and the different energies only after approximately 150 Myr is that, up to that point, the gas has not yet cooled enough for the intermediate- and high-density gas to be present in a stable state.
In the initial condition, $\langle\log\beta\rangle=6.3$, and the total thermal and magnetic energies are $10^{56.6}$ and $10^{54.4}$~~ergs, respectively.

In both models, the low-density gas presents high values of $\langle\log\beta\rangle$ indicating thermal dominance throughout the model evolution. However, there is a decrease of $\beta$ with time because the gas is cooling. The plasma $\beta$ in Model R  
is slightly lower than in model T, which is expected since we showed that the magnetic field of model R can be strong in regions with low-density gas (see Fig. ~\ref{fig:col_den}). Calculating the temporal average and standard deviation, we find  
$\rm \langle log~\beta\rangle_T =3.1 \pm 1.8$ 
which is an order of magnitude higher than the temporaly averaged value of 
$\beta$ in the low-density gas of model R, 
$\rm \langle log~\beta\rangle_R =2.3 \pm 2.1$. 

The intermediate-density gas starts in a state of thermal dominance with a value of $\langle\log\beta\rangle$ around 0.6 in both models, which is much lower than the initial values of the low-density gas. As the models evolve, $\langle\log\beta\rangle$ reaches an equipartition value in model R, while in model T, it remains in the thermally dominant region, albeit close to equipartition. Calculating the temporal average and standard deviation, 
we find a higher value of $\beta$ in model T
$\rm \langle log~\beta\rangle_T =0.96 \pm 1.7$.
compared to model R 
($\rm \langle log~\beta\rangle_R =0.1 \pm 1.5$).

The high-density gas is present starting from 180 Myrs in model T, while in model R, it forms earlier, at 120 Myrs. In both models, $\langle\log\beta\rangle$ starts with lower values (-1.3 in both cases), indicating magnetic dominance. However, in model T, $\langle\log\beta\rangle$ increases over time, finally reaching equipartition, whereas in model R, it remains in the magnetically dominated region throughout the evolution. However, they exhibit similar temporally-averaged values, 
$\rm \langle log~\beta\rangle_T =-0.16 \pm 1.2$ for model T and $\rm \langle log~\beta\rangle_R =-0.72 \pm 1.1$ for model R.
We note the large scatter around all the above averages, which mainly comes from the large scatter in the individual $\beta$ values.
With this large scatter in mind, the average values indicate a stronger dynamical significance of the magnetic with respect to thermal energy in model R compared to model T.

It is indeed crucial to consider the high standard deviation in the data, as it indicates significant variability within the gas included in each density phase. The temperature two-dimensional (2D) histogram shown in the bottom plot of Fig.~\ref{fig:beta} illustrates that the medium in the same density range exhibits different temperatures, leading to diverse distributions of log($\beta$) that can exhibit multiple peaks. This variability should be taken into account when analyzing and interpreting the results.

In the bottom panel of Fig.~\ref{fig:beta_t}, we show the total thermal (circles) and magnetic (triangles) energy of the low-density (blue), the intermediate-density (orange), and the high-density gas (green). 
In the initial conditions, the total thermal energy of the low-density gas in both models is two orders of magnitude higher than its total magnetic energy. However, by 500 Myr, these energies tend to become equal in model T. On the other hand, in model R, the total magnetic energy eventually surpasses the thermal energy.
For both models, the total magnetic energy of the intermediate and high-density gas remains higher than the thermal energy throughout the entire evolution, except for the time period before 100 Myr, where the energies are approximately equal in the intermediate phase.

It is important to note that the $\langle\log\beta\rangle$ values do not perfectly match the total energies due to several factors. Firstly, the ratio of total energies does not directly translate into the mean of log($\beta$), as they represent different physical quantities and have distinct interpretations.
Moreover, the high standard deviation observed in the data contributes to the tension. The variability in log($\beta$) within the density ranges can lead to a wide spread of values, affecting the mean and making it less representative of the entire gas population.

We also present one-dimensional (1D) radial profiles of thermal and magnetic energies for the three different gas phases at three different 
times (200, 300, and 500 Myrs) in App.~\ref{appendix:b}. 
These profiles also illustrate that, 
model R consistently exhibits a higher
degree of magnetic dominance 
than model T across different densities, and over time. 
They also show a slightly wider distribution of dense gas in model R compared to model T.

\subsection{\rm B-$\rho$ relations}\label{ssec:B_rho}

\begin{figure*}
    \centering
    \hspace{-.6cm} Model T  \hspace{6.2cm} Model R \hspace{4.5cm}\\
    \includegraphics[trim={0cm 0.cm 5.1cm 0.5cm},clip,width=.45\linewidth]{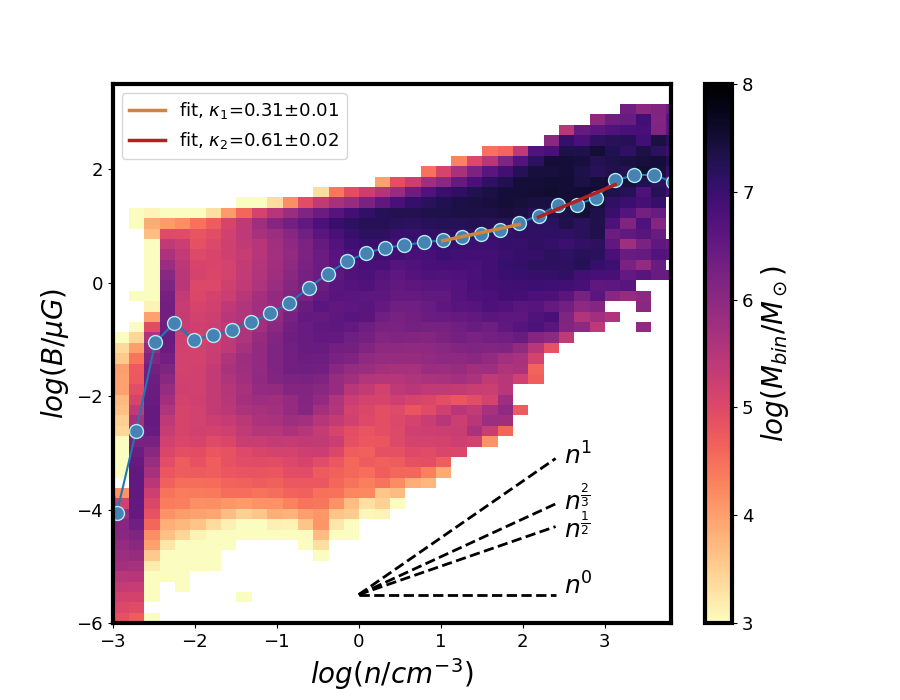} 
    \includegraphics[trim={1.72cm 0.cm 2.cm 0.5cm},clip,width=.48\linewidth]{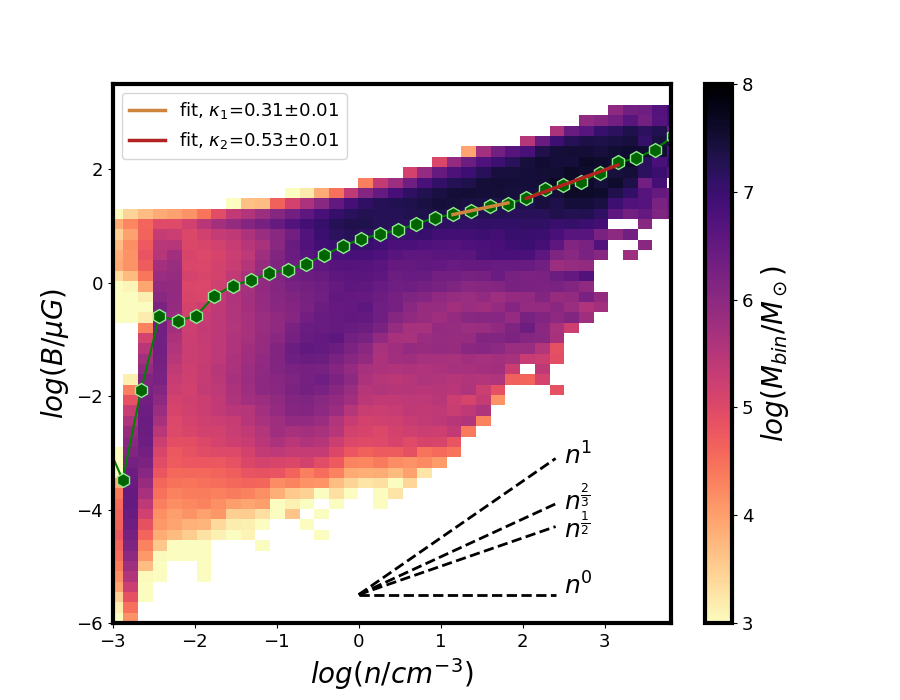}
    \caption{|B| versus {\bf n} for the two models at t = 470 Myr with the model T (left-hand side) and the model R (right-hand side).
    The B-$\rho$ relations are given as 2D mass-weighted histograms in the log B-log {\bf n} space.
    The dots indicate the mean magnetic field strength per density bin.
    The yellow and red lines are the power-law fits with slopes given in the legend of each plot. The B-$\rho$ relations derived from \cite{mestel,Mousc1,Mouschovias2,tritsis15} are reported with dashed black lines.
    \label{fig:b_rho}
    }
\end{figure*}

Fig.~\ref{fig:b_rho} shows the magnetic field-number density (B-n) relation at t=470 Myr. 
Model T is on the left-hand side and the model R is on the right-hand side. As in Fig.~\ref{fig:beta} and ~\ref{fig:beta_t}, we take into account only the gas cells in the disk, with $\rm|z|\leq1.5~kpc$ and radius $\rm R\leq13~kpc$.
The B-n relations are given as 2D mass-weighted histograms in the log B-log n space. The colored dots indicate the mean magnetic field strength per density bin. The standard error of the mean is not noticeable here because it is very small. We fit the mean relation using a broken power law: 
\begin{equation}\label{eq:broken_power_law}
\rm log (B/\mu G) =
    \begin{cases}
      \rm\kappa_1~log (n/cm^{-3}) & \rm n < n_0 \\
      \rm\kappa_2~log (n/cm^{-3}) & \rm n > n_0
    \end{cases}       ,
\end{equation}
where the two power-law slopes $\rm \kappa_1$ and $\rm \kappa_2$ are parameters of the fit, while the transition number density, $\rm n_0$, is fixed in this instance to 100 $\rm cm^{-3}$.  
In this figure we show only one snapshot to showcase the break, but later in this section we will examine the impact of varying the parameter $\rm n_0$ and show the time evolution of the slopes.
The broken power-law fit was chosen to test the suggestion that the B-$\rho$ relation should be flat below, and a power-law above, a critical density $n_0$ \citep{crutcher99,pattle}. However, even a simple visual inspection of Fig.~\ref{fig:b_rho} indicates that a single power law is a better fit to the B-$\rho$ relations in both models. Therefore, in the following we will also present results for a single power-law fit with an index simply referred to as $\kappa$.

The slopes $\rm \kappa_1$ and $\rm \kappa_2$ are provided in the legend of each plot. Additionally, in the bottom right corner of each plot, we include the B-$\rho$ relations proposed by various theories, which are based on different assumptions regarding conservation laws and geometry during the collapse (as discussed in Section \ref{sec:introduction}).
We recall here that $B \propto n^0$ results from compression along magnetic field lines, $B \propto n^1$ from compression perpendicular to the magnetic field lines, $B \propto n^{\frac{1}{2}}$ implies a slab-like or filamentary geometry, where the magnetic field lines are either perpendicular to the slab or inclined relative to the primary axis of the filament, which collapses radially and $B \propto n^{\frac{2}{3}}$ arises from spherical compression \citep[see Fig. 1 from][for a clear illustration]{tritsis15}. 

In both models, there is a positive correlation between the magnetic field and density. In the intermediate- and high-density gas phase, we do not observe the flatness in the magnetic fields reported in observations for $\rm n<300~cm^{3}$ \citep{crut10, pattle}.
However, when we focus on the peak of the mass, represented by darker regions, at diffuse medium densities we observe a bimodal distribution. One branch appears to extend the high-density power law to lower densities, while the other branch shows a flattening of the magnetic field close to 10 $\mu$G. This behavior has been consistent across all the times we have examined.

Comparing the two panels, for both models we see that the B-n relation at intermediate densities has a power-law slope of $\rm \kappa_1=0.31 \pm 0.01$. For both models, in the high-density medium, the power law steepens with respect to the intermediate, with $\rm \kappa_2=0.61 \pm 0.02$ for model T and $\kappa_2=0.53 \pm 0.01$ for model R.
According to \cite{mestel,Mousc1,Mouschovias2,tritsis15}, the dense gas behavior of model T is closer to the theoretical prediction for isotropic spherical compression while the behavior of model R implies a slab or filament-like geometry.

\begin{figure*}
    \centering
    \rotatebox{90}{\hspace{2.0cm} $n_0=50\; cm^{-3}$}
    \includegraphics[trim={0cm 0cm 0cm 0cm},clip,width=.4\linewidth]{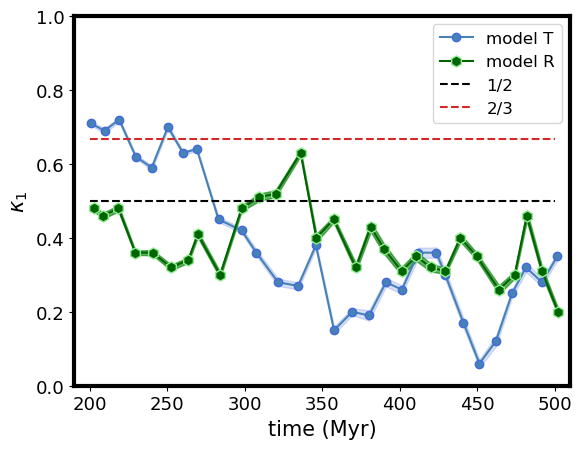} 
    \includegraphics[trim={0cm 0cm 0cm 0cm},clip,width=.4\linewidth]{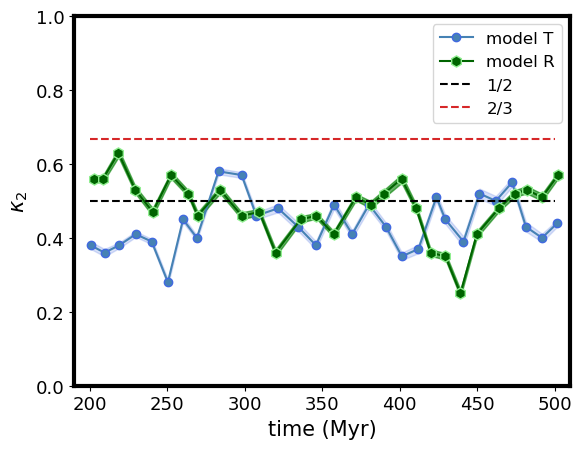}\\
    \rotatebox{90}{\hspace{2.0cm} $n_0=100\; cm^{-3}$}
    \includegraphics[trim={0cm 0cm 0cm 0cm},clip,width=.4\linewidth]{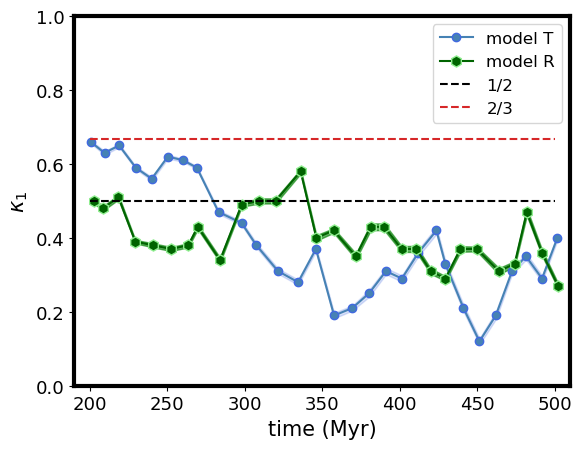} 
    \includegraphics[trim={0cm 0cm 0cm 0cm},clip,width=.4\linewidth]{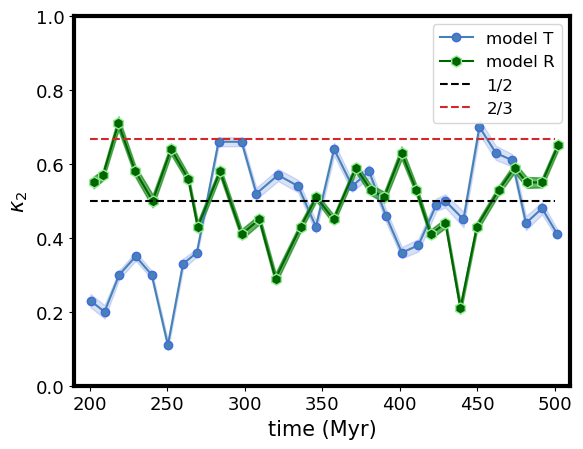}\\
    \rotatebox{90}{\hspace{2.0cm} $n_0=300\; cm^{-3}$}
    \includegraphics[trim={0cm 0cm 0cm 0cm},clip,width=.4\linewidth]{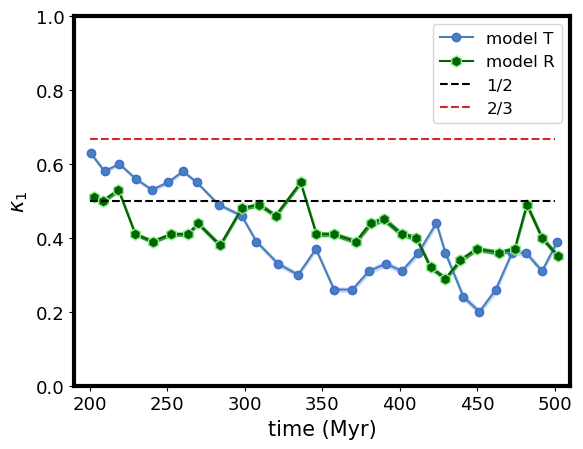} 
    \includegraphics[trim={0cm 0cm 0cm 0cm},clip,width=.4\linewidth]{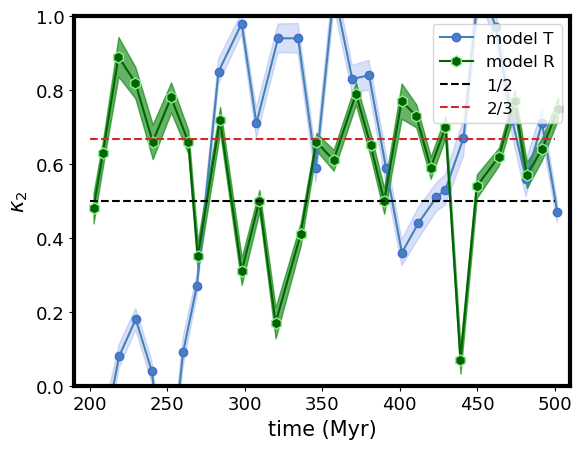} \\
    \rotatebox{90}{\hspace{2.0cm} Single power-law}
    \centering
    \includegraphics[trim={0cm 0cm 0cm 0cm},clip,width=.4\linewidth]{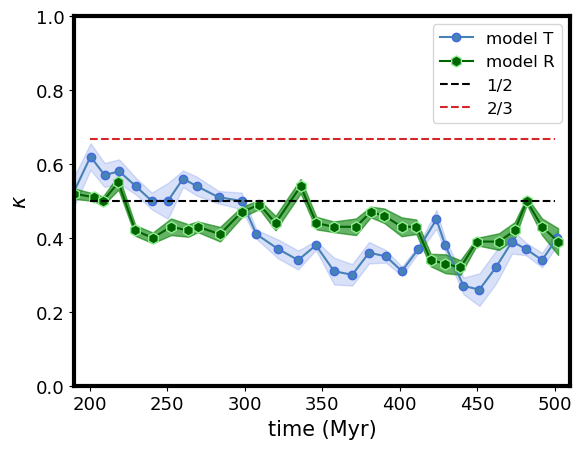}
    \caption{Evolution of the $B-\rho$ slopes ($\rm \kappa_1, \kappa_2$) and the slope $\kappa$ from the single power-law fit as a function of time for model T (blue) and model R (green) for intermediate (left-hand side) and high densities (right-hand side).
    The different transition densities for the broken power-law fit are indicated as labels on the left. 
    The definition of the broken power law fit is given in Eq. \ref{eq:broken_power_law}.
    The shaded areas correspond to the standard deviation error on the slopes derived from the fit. The black and red dashed lines show the slopes of 1/2 and 2/3 respectively.}
    \label{fig:b_rho_t}
\end{figure*}

However, it is crucial to highlight that this is for a single snapshot in time and that the unique aspect of model R does not follow the conventional assumption of a uniform magnetic field. Therefore, the existing theoretical predictions may not directly capture the specific dynamics and behaviors observed in model R. 
Using the same fitting procedure for the B-n relation, in Fig.~\ref{fig:b_rho_t} we plot the slopes ($\kappa_1, \kappa_2$) for the broken power-law fit and the slope $\kappa$ for the single power-law fit as a function of time for Model T (blue) and Model R (green).

In order to examine the dependence of the fitted exponents on the value of the transition density we consider three different values: $\rm n_0=50\,\cc$, shown in the top panel, $\rm n_0=100\,\cc$, shown in the middle panel, and $\rm n_0=300\,\cc$, shown in the bottom panel. We chose the particular transition densities because according to \cite{crut10} we expect the break of the power law at $\rm n_0=300\,\cc$. However, we investigate additional density ranges, as the break may occur at different values depending on time.
The shaded areas correspond to the standard deviation (1-$\rm \sigma$) of the slopes derived from the fit. The black and red dashed lines represent the slopes of 1/2 and 2/3, respectively. It is worth noting that both slopes vary significantly over time.

In Model T and for a transition density of $\rm n_0=50\; cm^{-3}$, the intermediate-density slope $\rm \kappa_1$ fluctuates around 2/3 from 200 to 300 Myr. Later on, it decreases to values lower than 1/2, indicating compression predominantly along the magnetic field lines. Instead, in model R, $\kappa_1$ fluctuates around 0.4 for the entire duration of the simulation. 
Eventually, both models converge to similar values around 0.3-0.4. For higher transition densities ($\rm n_0=100\,\cc$ and $\rm n_0=300\,\cc$), the behavior of $\rm \kappa_1$ is similar to that of $\rm n_0=50\,\cc$, but the variations are smoother.

For model R, and a transition density of $\rm n_0=50\,\cc$, the high-density slope $\kappa_2$ fluctuates around 1/2, implying a slab or filament-like geometry with a perpendicular or radial compression. 
At the same transition density, $\kappa_2$ for model T is slightly lower, staying slightly below 1/2 for almost the entire duration of the simulation. 
Calculating the temporal average of the slopes for $n_0=50\,\cc$, we find $\langle\kappa_1\rangle_T = 0.34 \pm 0.19$ for model T and $\langle\kappa_1\rangle_R =0.38 \pm 0.09$ for model R. Model T also exhibits $\langle\kappa_2\rangle_T= 0.44 \pm 0.07$, and model R records an average slope of $\langle\kappa_2\rangle_R=0.48 \pm 0.08$. The errors are attributed to the time evolution which is the primary source of variance.
For higher $\rm n_0$ values, $\kappa_2$ exhibits significant variations and differences between the two models. 

For a transition density of $\rm n_0=100\,\cc$, $\kappa_2$ starts off near 0.2 for model T and 0.6 for model R. After about 250 Myr and for the remainder of the simulation, $\kappa_2$ in both models R and T fluctuates around 1/2 but with larger amplitude (between 0.4 and 2/3) than for the lower transition density.
Calculating the temporal average of the slopes, $\langle\kappa_1\rangle_T =0.36 \pm 0.16$ for Model T and $\langle\kappa_1\rangle_R =0.40 \pm 0.07$ for Model R. For higher densities ($>n_0$), Model T exhibits a slope of $\langle\kappa_2\rangle_T=0.46 \pm 0.15$, and Model R $\langle\kappa_2\rangle_R =0.51 \pm 0.10$.

In the case of $\rm n_0=300\,\cc$, the initial values of $\rm \kappa_2$ for Model T are close to zero, indicating compression along magnetic field lines, which is the theoretical expectation for an ordered field morphology initially. After 280 Myr, $\rm \kappa_2$ in model T shifts to unity, the theoretical expectation for compression across the field lines. At later times, $\rm \kappa_2$ fluctuates strongly between 0.5 and 1. 
Also for R $\rm \kappa_2$ fluctuates strongly, although not following any of the theoretical expectations. The stronger fluctuations as the threshold density increases are due to the smaller fitting range, but also due to poorer statistics, because at this density range, we are occasionally depleting gas to star formation.

Calculating again the temporal average, the slopes for intermediate densities are $\langle\kappa_1\rangle_T =0.38 \pm 0.12$ for Model T and $\langle\kappa_1\rangle_R=0.42 \pm 0.06$ for Model R. In the case of high densities, Model T shows a slope of $\langle\kappa_2\rangle_T=0.53 \pm 0.39$, while Model R $\langle\kappa_2\rangle_R=0.60 \pm 0.19$.

Overall, the $\rm B-\rho$ relation exhibits strong variations across different density ranges and over time. However, despite these variations, the average intermediate slope ($\langle\kappa_1\rangle$) consistently appears lower than the average high-density slope ( $\langle\kappa_2\rangle$). Additionally, for higher densities, both slopes tend to converge toward 1/2 for lower transition densities ($n_0$) which gives better statistics. This consistency aligns with previous studies by \cite{Mousc1,Mouschovias2}.

The single power-law fit (bottom row of Fig. \ref{fig:b_rho_t}) is very similar in temporal behavior to that of $\kappa_1$. The slopes in the two models both start near $\kappa$=0.5 and decrease over time, to a value close to $\kappa$=0.4. While there are time variations of the same order as what we noted for $\kappa_1$ in the previous cases, when averaged over time, the $\kappa$ values in the two models are within one sigma from each other.
Finally, it is worth noting that the $\chi^2$ for the single power-law fit is 
lower than that of the double power law, for both models. We calculated the BIC (Bayesian Information Criterion, \citealt{schwarz1978}) for the two models as an indication for the preferred fit. The BIC:
\begin{equation}
    BIC = k~ln(n) - 2~ln(L)
\end{equation}
where k the number of parameters of the model, n the number of data points, and L the likelihood function, penalizes models with more fitting parameters, with lower values of BIC indicating a better fit.
Here as L we have taken the value of $\chi^2$). The BIC comparison favors the single power-law fit with respect to the broken power law for all snapshots, despite the small difference in the number of parameters between the two models.

\section{Summary and discussion}\label{sec:discuss}

\subsection{Summary}

In this work, we presented galaxy-scale MHD simulations that incorporate gravity, star formation, supernova feedback, and chemistry. For the first time, we explored different magnetic field morphologies, specifically an ordered toroidal magnetic field (model T) and a random magnetic field (model R). Our study focuses on star formation, the plasma beta parameter, and the $\rm B-\rho$ relation in the two models.

We observe that the star formation rates (SFRs) of the two models peak at different times, and there are some differences in the instantaneous SFR at different instants. However, the time-averaged SFR is similar between the two models, especially when taking into account the large scatter around the temporal mean. They also produce the same stellar mass after 500 Myrs of evolution. This implies that a different magnetic field morphology may result in a slightly different evolution of the SFR due to the complex nature of the problem, but this behavior does not affect the average star formation history of the galaxy.

We found that in both models, the low-density gas is thermally dominated, with model R showing slightly lower $\langle\log\beta\rangle$ values. The intermediate-density gas tends \textbf{toward} equipartition in model R but remains thermally dominant in model T, while the high-density gas forms earlier in model R and exhibits magnetic dominance throughout its evolution. Concerning the total thermal and magnetic energies, initially, the total thermal energy of low-density gas significantly exceeds its total magnetic energy in both models. However, by 500 Myr, these energies approach equality in model T, whereas in model R, the total magnetic energy eventually surpasses the thermal energy. Throughout the evolution, both models maintain higher total magnetic energy for the intermediate and high-density gas.
However, it is important to note that the scatter around the average $\beta$ values is always very large, so the above differences are within the errors.
Regarding the $\rm B-\rho$ relation, we observe a scaling that follows a power law with exponents $\kappa$ ranging from 0.2 to 0.8 and varying over time. The temporal evolution of $\kappa$ differs between the two models, but these disparities fall within the wide temporal scatter.

\subsection{Comparison with previous studies and related works}

Previous computational works have studied the $\rm B-\rho$ relation using different approaches and assumptions.

\cite{pardi} focused on kpc-sized portions of a galactic disk, examining the chemical, thermal, and dynamical evolution of the gas with self-gravity and supernova feedback. They found a large scatter around the $\rm B-\rho$ relation due to the dominance of kinetic energy density in their simulations. Unlike our results, they did not find a clear increase in magnetic field strength with density across the entire density range, suggesting a possible flattening of the relation above a density of $\rm 0.6\,\cc$. This flattening might be a result of a limited dynamical range.

Using a similar setup, \cite{Girichidis} found that weakly magnetized low-density gas ($\rm n<0.6\,\cc$) exhibited a power-law scaling with a slope of $\rm \kappa=2/3$, while higher densities ($\rm n>0.6\,\cc$) showed a flatter slope of $\rm \kappa=1/4$. Our results do not align with these studies since we observe scaling for densities higher than 10 $\rm cm^{-3}$, and moreover, the power law becomes steeper at higher densities. However, there are some differences in the setups between our study and the previous ones by \cite{pardi,Girichidis}. Notably, the previous studies focus on smaller-scale regions, do not simulate entire galaxies, and utilize initially uniform magnetic fields. Additionally, variations in the heating and cooling mechanisms as well as the star formation recipe employed could contribute to the differences observed. 

\cite{Seta} modeled the turbulent dynamo in a two-phase medium considering solenoidal and compressive driving. They initialized the simulation with a weak random seed field and explored a density range of $10^{-2}$ to $10^{4}\; \rm cm^{-3}$, which varied depending on the turbulence driving and temperature. They find that the slope of the $\rm B-\rho$ relation is shallower in the solenoidal ($\rm \kappa=0.22$ for $T<10^3\,\rm K$, $\rm \kappa=0.27$ for $T>10^3\,\rm K$ ) than in the compressive case ($\rm \kappa=0.71$ for $T<10^3\,\rm K$, $\rm \kappa=0.51$ for $T>10^3\,\rm K$ ). In our study, we focus on fitting the $\rm B-\rho$ relation for temperatures lower than $10^4\,\rm K$. 
Within the uncertainties due to the temporal variations, our models lie between these two cases.
This fact is not surprising because turbulence in our simulation is injected both from supernova events,  
which inject a mixture of compressive and solenoidal modes \citep{Pan2016}, 
and the differential rotation of the disk, which preferentially injects solenoidal modes. 
However, a simple calculation indicates that, in our models, the dominant contribution to turbulence comes from the differential rotation. The number of supernova explosions can be estimated from the number of stars formed per unit time, indicatively with a $\langle SFR\rangle\simeq 1~ M_{\odot}/yr$. With $20\%$ of this mass exploding as SNe, $\rm E_{SN}=10^{51}~erg$ the energy per SN and $25\%$ of $\rm E_{SN}$ going to kinetic energy injected in the ISM, we estimate the rate of turbulent kinetic energy from SNe to be approximately $E_{turb,SN}\simeq10^{49}$ erg/yr.
For the energy contribution from the differential rotation, we calculate the rotational velocity for each cell and use the relation $\rm K_{rot} = 1/2\,m\,v_{rot}^2$ to calculate the kinetic energy. We then divide this energy by the timestep, which is $10^3$ yr, to obtain an estimate of $E_{turb,rot}\simeq10^{54}$ erg/yr. It is important to note that cooling effects are not included in this calculation, and cooling could counteract some of the energy injection from supernovae.

Finally, in collapse simulations of an isothermal turbulent medium, \cite{brandenburg_ntormousi} employed both a random and a guide field with low and high magnetic field strengths. They consistently observed a large scatter around the relation, such that both $\kappa=2/3$ and $\kappa=1/2$ could fit the data. This finding is consistent with our study, where we also observe significant scattering in the relation.

\subsection{The impact on galactic physics and observations}

Variations in the $\rm B-\rho$ relation with time and with magnetic field morphology could have significant implications for our interpretation of the dynamical state of the gas and the star formation process.
Our analysis shows that the $\rm B-\rho$ relation is not universal but rather context-dependent, reflecting the interplay between magnetic fields and the 
evolutionary stage and magnetic field morphology of the observed galaxy.

In principle, a galaxy's magnetic field will change morphology as the galaxy evolves. Mergers and accretion, which are particularly relevant for early galaxies \citep[e.g.][]{kohandel:2020}, drive turbulence, which in turn can generate a turbulent dynamo. This would lead to a more random magnetic field morphology. Instead, a mean-field dynamo could take over in more quiescent phases, leading to more coherent large-scale fields \citep[e.g.,][for a review on galactic dynamos]{BN2022review}.
However, our results imply that any influence of the galactic magnetic field's morphology on the B-$\rho$ relation must be smaller than the scatter introduced by the non-linear processes in the ISM, since the differences between the two models always fall within a very broad temporal scatter.
However, we should stress that our experiments do not follow the complex, long-term evolution of the magnetic field from primordial seeds to the present day. Both models contain strong initial fields and none of them shows dynamo action. The imprints of the complex co-evolution of the galaxy and its magnetic field will be the subject of future work.
Our results also suggest that
interpreting the $\rm B-\rho$ relation in observations should be done cautiously, given the large scatter and the complex dependencies discovered in our models. 
While our study acknowledges that we are currently lacking proper production of observable-like data, it still emphasizes the importance of considering various factors and specific galactic conditions in comprehensively understanding the role of magnetic fields in star formation.

\subsection{Future improvements: early feedback}

One of the innovations in our work is the inclusion of non-equilibrium chemistry in the simulations through the KROME package, allowing us to follow the formation and dissociation of $\rm H_2$. This leads to a more realistic representation of star formation
because
it is based on the 
$\rm H_2$ fraction. However, it is important to note that our simulations currently lack pre-SN feedback. Including early feedback, namely radiation and stellar winds alter their impact \citep{pallo,pallo_19,decataldo}, in particular increases the time variability of the SFR \citep{pallottini:2023}, which in turn might enhance the scatter in the $B-\rho$ relation for some gas phases. 
The inclusion of early feedback is clearly the next step for this study.

\section{Conclusions}\label{sec:conclus}

Our analysis of galaxy-scale MHD simulations with different magnetic field morphologies contributes to the understanding of the $\rm B-\rho$ relation as a probe of the ISM's dynamical state with several important findings:

\begin{enumerate}
    \item The two models (with ordered and random magnetic field) have a similar average star formation rate and they end up forming roughly the same total mass in stars, implying that the evolution of galaxies with comparable formation histories can have unique ISM characteristics depending on the magnetic field morphology.
    \item The plasma beta values indicate different thermal and magnetic dominance behaviors across different gas density phases. Both models have thermally dominated low-density gas ($\rm 10^{-2} < n/cm^{-3} < \rm 1$) with $\langle\log\beta\rangle$>3, but model R shows slightly lower $\langle\log\beta\rangle$ values. In the intermediate-density gas ($\rm 1 < n/cm^{-3} < \rm 10^{2}$), model R tends toward equipartition, while model T remains thermally dominant with $\langle\log\beta\rangle \approx 1$. For high-density gas ($\rm 10^{2} < n/cm^{-3} < \rm 10^{4}$), model R exhibits magnetic dominance throughout its evolution with $\langle\log\beta\rangle \approx -1$, whereas in model T high-density gas forms later and reaches equipartition.
    \item The total magnetic energy remains higher than the thermal energy for the intermediate and high-density gas phases in both models throughout their evolution. However, in the case of low-density gas, the total thermal energy initially exceeds the total magnetic energy for both models. By 500 Myr, these energies approach equality in model T as the galaxy cools down and the magnetic field strength slightly increases. On the other hand, in model R, the total magnetic energy eventually surpasses the thermal energy due to a slightly stronger magnetic field compared to model T.
    \item The $\rm B-\rho$ relation is not universal even within the same galaxy, displaying bimodal distributions and significant variations over time with $\kappa$ ranging from about 0.2 to 0.8. Despite the variations, the slope $\kappa$ at densities $\rm n>50,\cc$ tends to converge to 1/2 for both models.
    \item Even considering the large scatter, a single power-law fit describes the B-$\rho$ relation at intermediate and high densities ($1<n/cm^{-3}<10^3$) better than a broken power law, independently of the choice of the transition density, $n_0$.
    \item The observed differences in the slope evolution between the two models are not large enough to indicate that the magnetic field morphology influences a galaxy's $\rm B-\rho$ relation because they fall within the very large scatter.
\end{enumerate}

Overall, we see a small influence of the magnetic field morphology on a galaxy's $\rm B-\rho$ relation, which is, however, transient and below the level of the relation's fluctuations due to other stochastic phenomena. Eventually, both models tend to $B\propto \rho^{0.5}$. 
In general, the  differences in between the two models, while measurable at each individual time, show significant scatter over time, highlighting the complex and context-dependent nature of the $\rm B-\rho$ relation.
In closing, we should emphasize that, due to limited resources, our study only included two models of gas-poor, massive spiral galaxies. The effects of the magnetic morphology on the gas dynamics could be very different for gas-rich, more turbulent, or less massive galaxies.
Further research and comprehensive observations will be crucial to fully understand the intricate role of magnetic fields in star formation processes across diverse galactic environments.

\begin{acknowledgements}
    We would like to thank A. Ferrara, P. Hennebelle, R. Teyssier, T. Ch. Mouschovias, R. Skalidis, N. Loudas, G. Korkidis, and K. Kovlakas for insightful discussions related to this project.
    This project has received funding from the European Research Council (ERC) under the European Unions Horizon 2020 research and innovation programme under grant agreement No. 771282 and from the Hellenic Foundation for Research and Innovation (HFRI), second call for post-doctoral researchers, project number 224.
    This work was supported by computational time granted from the National Infrastructures for Research and Technology S.A. (GRNET S.A.) in the National HPC facility - ARIS - under project ID pr013007$\_$thin. 
    Part of the work has been performed under the Project HPC-EUROPA3 (INFRAIA-2016-1-730897), with the support of the EC Research Innovation Action under the H2020 Programme.
    The authors gratefully acknowledge the support of Scuola Normale Superiore di Pisa (SNS) and the computer resources and technical support provided by the CINECA supercomputing center.
    We also gratefully acknowledge the computational resources of the Center for High Performance Computing (CHPC) at SNS and at the University of Crete.
    We acknowledge usage of the Python programming language \citep{python2,python3}, Matplotlib \citep{matplotlib}, NumPy \citep{numpy}, and SciPy \citep{scipy}.
\end{acknowledgements}


\bibliographystyle{aa}
\bibliography{citations}

\begin{appendix}

\onecolumn
\section{Evolution of thermal and magnetic phase diagrams}\label{appendix:a}

\begin{figure*}[!ht]
    \centering
    \hspace{-1.cm} Model T  \hspace{4.1cm} Model R \hspace{8.cm}\\
    \rotatebox{90}{\hspace{1.7cm} t = 200 Myr}
    \includegraphics[trim={0.45cm 1.8cm 5.2cm 2.cm},clip,width=.36\linewidth]{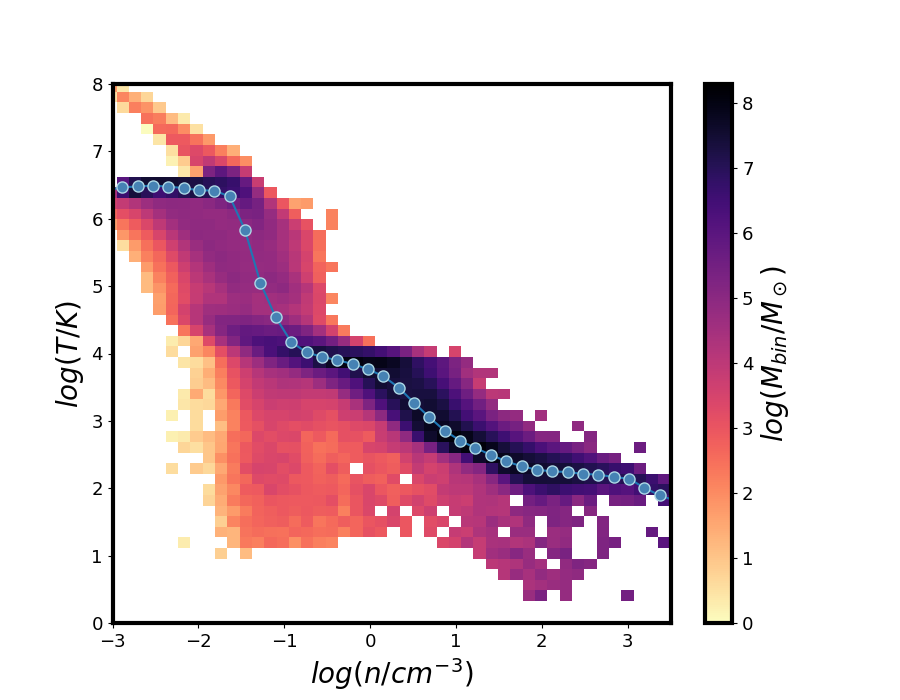} 
    \includegraphics[trim={2.75cm 1.8cm 0cm 2.cm},clip,width=.421\linewidth]{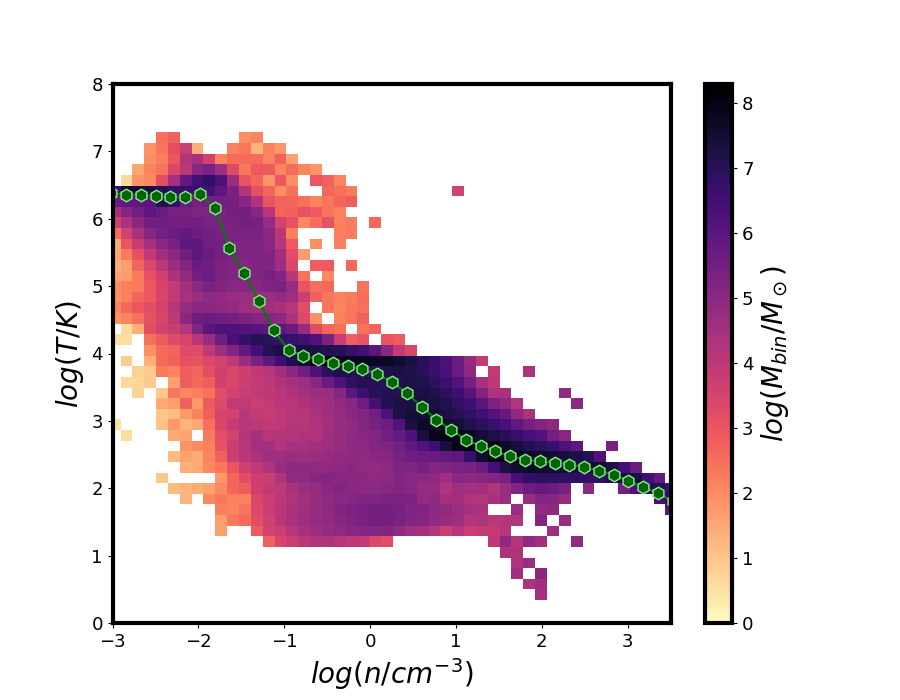}\\
     \rotatebox{90}{\hspace{1.7cm} t = 300 Myr}
    \includegraphics[trim={0.45cm 1.8cm 5.2cm 2.cm},clip,width=.36\linewidth]{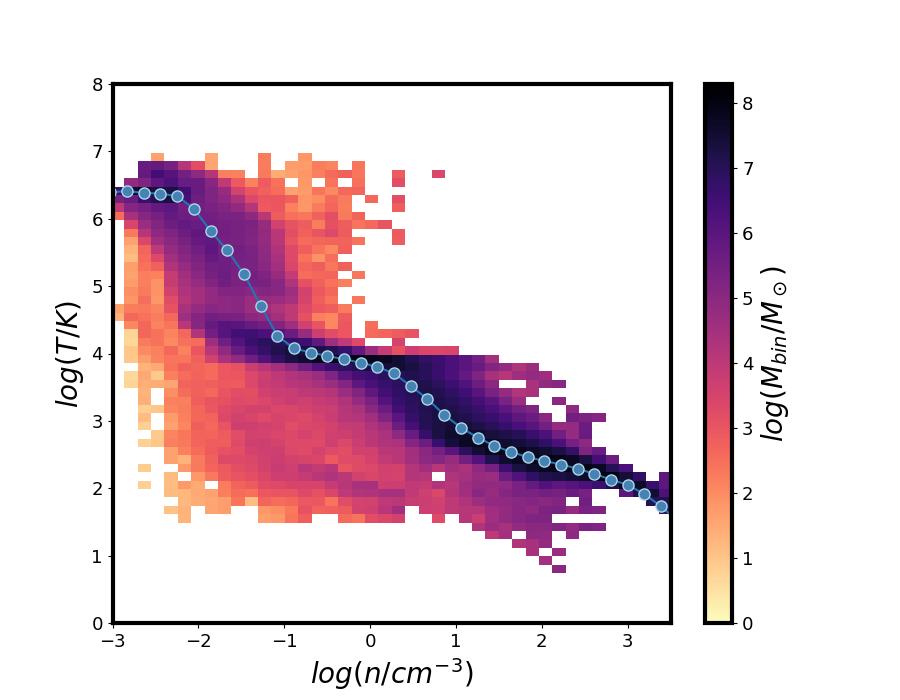} 
    \includegraphics[trim={2.75cm 1.8cm 0cm 2.cm},clip,width=.421\linewidth]{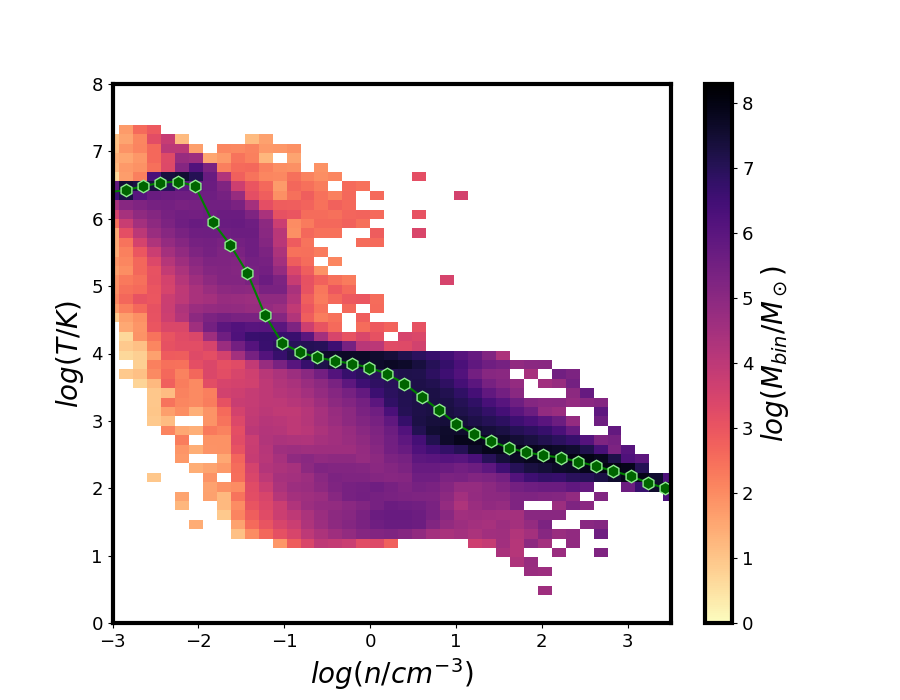}\\
    \rotatebox{90}{\hspace{2.4cm} t = 500 Myr}
    \includegraphics[trim={0.45cm 0.cm 5.2cm 2.cm},clip,width=.36\linewidth]{figures/T_rho_2d_hist_500.0.png} 
    \includegraphics[trim={2.75cm 0.cm 0cm 2.cm},clip,width=.421\linewidth]{figures/T_rho_2d_hist_517.0.png}
    \caption{2D mass-weighted histograms of the logarithm of the temperature of the gas plotted against log(n) for model T (left-hand side) and model R (right-hand side) at t = 200, 300, and 500 Myr. The colored dots represent the mean temperature per density bin.}
    \label{fig:T_rho_t}
\end{figure*}

Fig.~\ref{fig:T_rho_t} illustrates 2D mass-weighted histograms presenting the logarithm of the gas temperature plotted against the logarithm of gas density. The left side of the figure represents model T, while the right side corresponds to model R. The histograms exhibit snapshots at different time points: t = 200, 300, and 500 Myr. Each histogram features colored dots indicating the mean gas temperature per density bin. Over time, model R develops a significant gas fraction at low densities ($10^{-1}<n/\,\cc<10\,\cc$) characterized by temperatures below 100K. In contrast, this specific temperature-density regime remains unoccupied in the corresponding plot for model T.

\begin{figure*}[!ht]
    \centering
    \hspace{-1.cm} Model T  \hspace{4.1cm} Model R \hspace{8.cm}\\
    \rotatebox{90}{\hspace{1.7cm} t = 200 Myr}
    \includegraphics[trim={0.45cm 1.8cm 5.2cm 2.cm},clip,width=.35\linewidth]{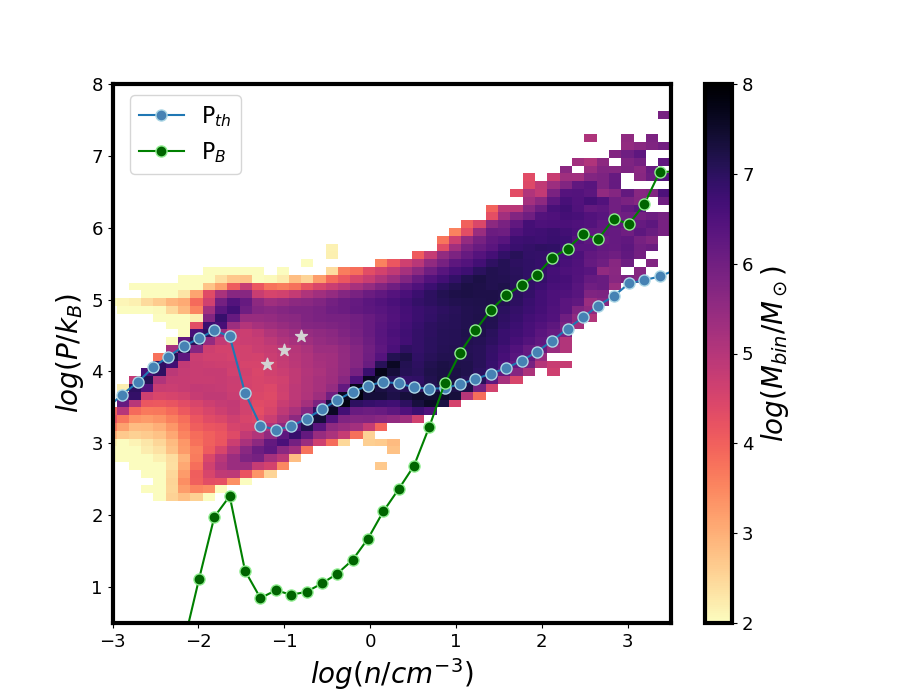} 
    \includegraphics[trim={2.75cm 1.8cm 0cm 2.cm},clip,width=.411\linewidth]{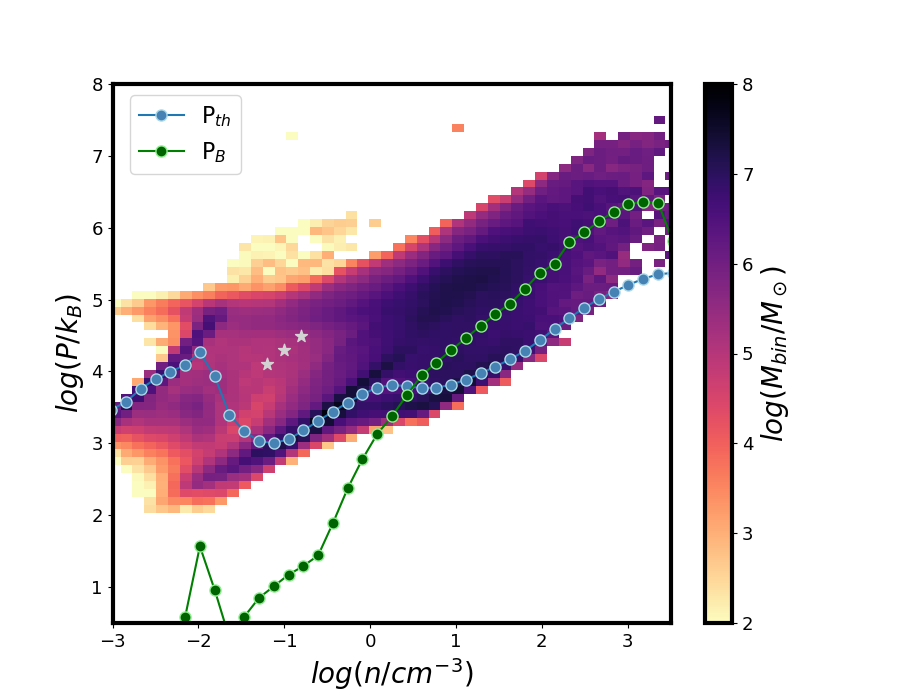}\\
    \rotatebox{90}{\hspace{1.7cm} t = 320 Myr}
    \includegraphics[trim={0.45cm 1.8cm 5.2cm 2.cm},clip,width=.35\linewidth]{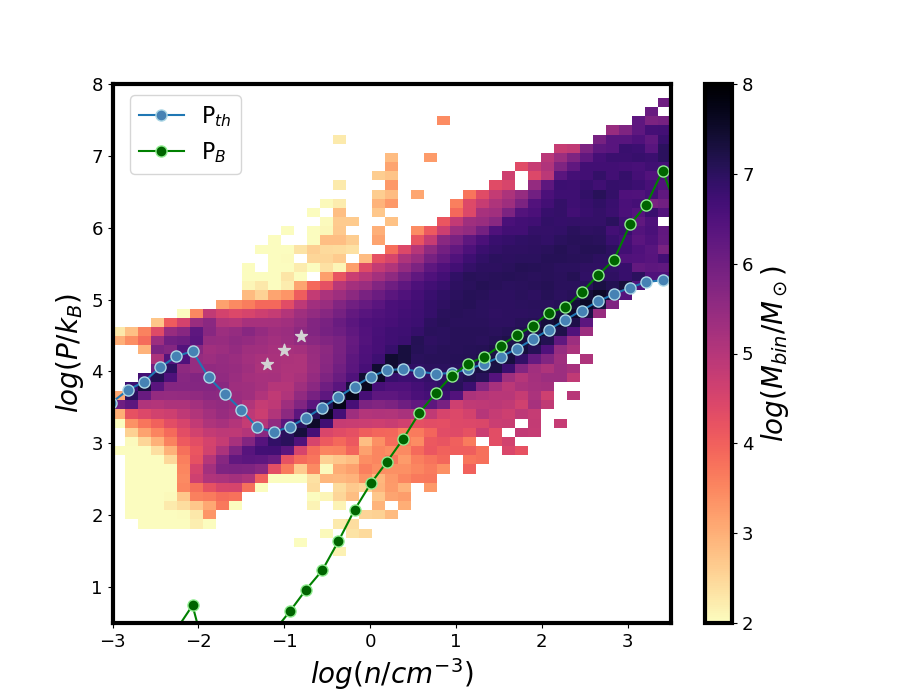} 
    \includegraphics[trim={2.75cm 1.8cm 0cm 2.cm},clip,width=.411\linewidth]{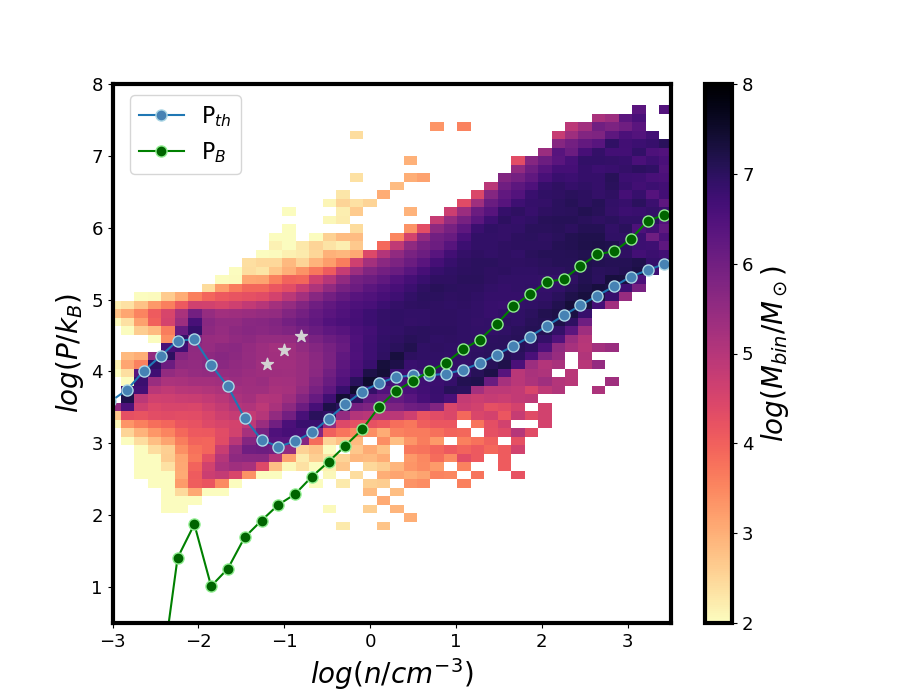}\\
    \rotatebox{90}{\hspace{1.7cm} t = 450 Myr}
    \includegraphics[trim={0.45cm 1.8cm 5.2cm 2.cm},clip,width=.35\linewidth]{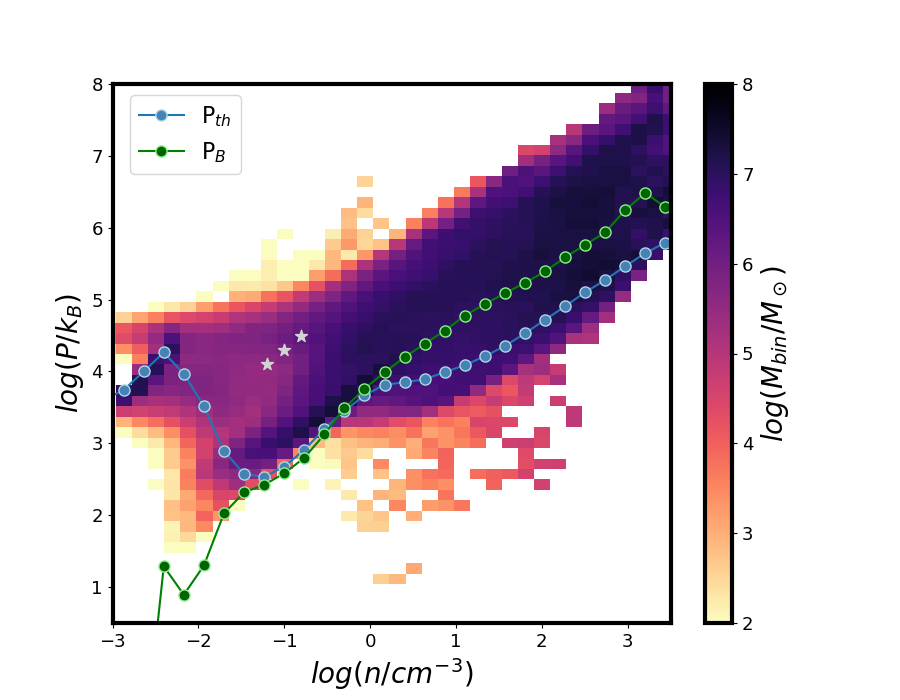} 
    \includegraphics[trim={2.75cm 1.8cm 0cm 2.cm},clip,width=.411\linewidth]{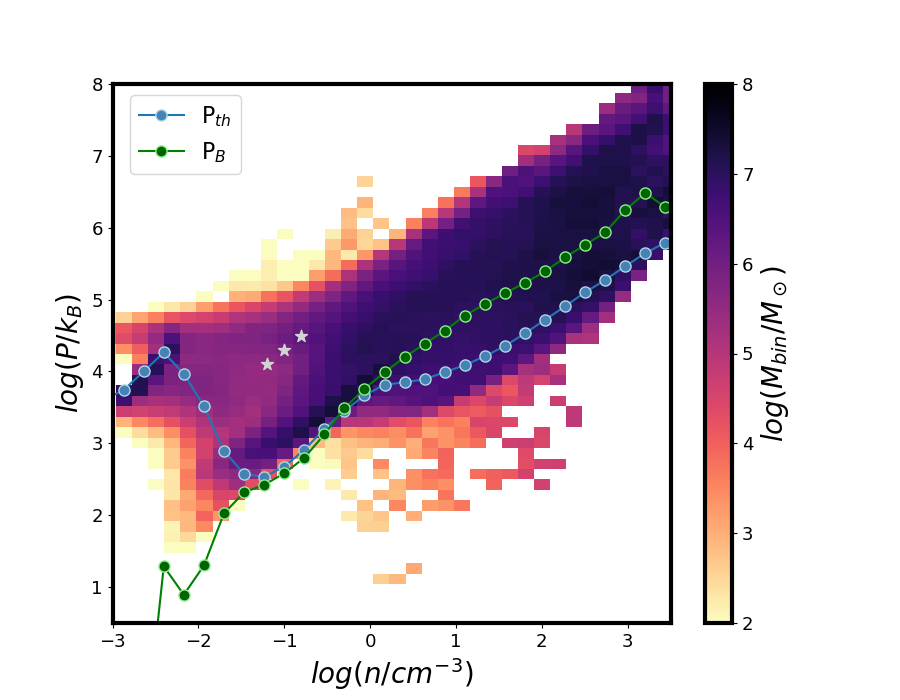}\\
    \rotatebox{90}{\hspace{2.4cm} t = 500 Myr}
    \includegraphics[trim={0.45cm 0.cm 5.2cm 2.cm},clip,width=.35\linewidth]{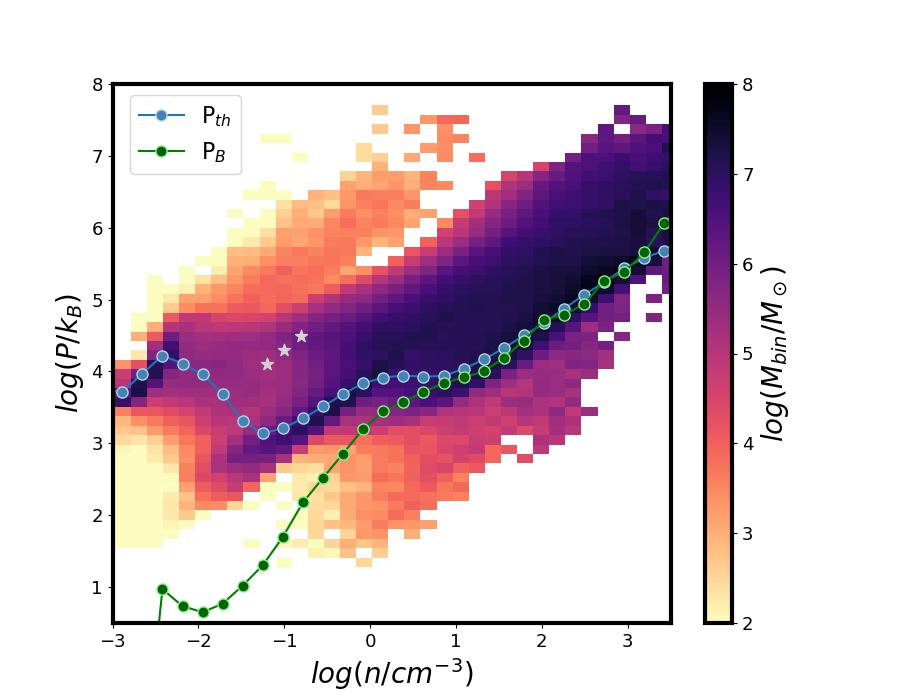} 
    \includegraphics[trim={2.75cm 0.cm 0cm 2.cm},clip,width=.411\linewidth]{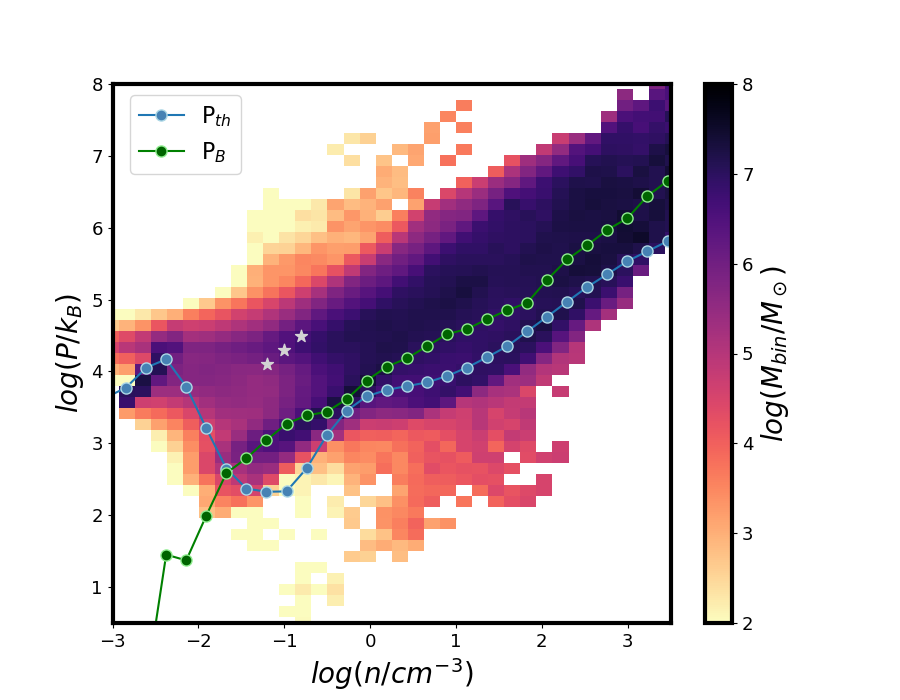}
    \caption{2D mass-weighted histograms of the logarithm of the total pressure of the gas plotted against log(n) for model T (left-hand side) and model R (right-hand side) at t = 200, 320, 450 and 500 Myr. The blue dots represent the mean thermal pressure per density bin while the green ones the magnetic pressure. The grey stars represent the initial condition.}
    \label{fig:p_n_t}
\end{figure*}

Fig.~\ref{fig:p_n_t} displays 2D mass-weighted histograms showing the logarithm of the total gas pressure plotted against the logarithm of gas density. The left side of the figure corresponds to model T, while the right side corresponds to model R. The histograms are shown at different time snapshots: t = 200, 320, 450, and 500 Myr. In each histogram, the blue dots represent the mean values of the thermal pressure per density bin, while the green dots represent the mean values of the magnetic pressure per density bin. The initial conditions are indicated by grey stars. 

At 200 Myr, both models are thermally dominated for densities lower than $\rm 10\,\cc$ and magnetically dominated for higher densities. As time progresses, model T transitions to equipartition, with magnetic and thermal contributions becoming comparable in higher densities, while maintaining thermal dominance in lower densities. In contrast, model R continues to become magnetically dominated even in lower densities over time.

\section{Radial profiles of thermal and magnetic energies for the three different phases of the gas}\label{appendix:b}

We plot 1D radial profiles of thermal and magnetic energies for the three different phases of the gas.
The initial condition for both models comprises exclusively low-density gas, where thermal energy dominates over magnetic energy across the entire range of radii. However, in the central region of the galaxy, the magnetic energy approaches  the thermal energy, and it rapidly decreases as we move to larger radii compared to the thermal energy.

At 200 Myrs, the low-density gas is thermally dominated across the entire range of radii for model T. In contrast, for model R, within the first 1 kpc from the center, the mean magnetic energy is slightly higher than the thermal energy.
The medium-density gas is magnetically dominated for the first 4 kpc from the center in both models. Then, it becomes thermally dominated. 
On the other hand, high-density gas extends up to 4.5 kpc from the center in model T, where it remains magnetically dominated. In Model R, it primarily extends to 8 kpc, with occasional occurrences as far as 12.5 kpc. In model R, the high-density gas is magnetically dominated until 5 kpc and then transits to thermal dominance.

\begin{figure*}[!ht]
    \centering
    t=200 Myrs
    \vspace{0.5cm}\\
    \hspace{1.0cm} Model T  \hspace{5.1cm} Model R \hspace{8.cm}\\
    \rotatebox{90}{\hspace{2.1cm} Low-density}
    \includegraphics[trim={0.cm 0.cm 0.cm 0.cm},clip,width=.4\linewidth]{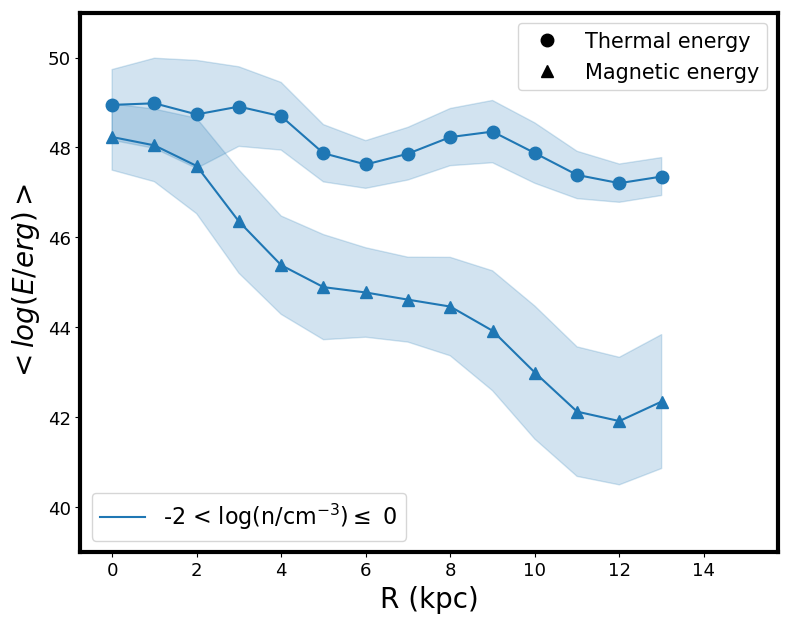} 
    \includegraphics[trim={0.cm 0.cm 0cm 0.cm},clip,width=.4\linewidth]{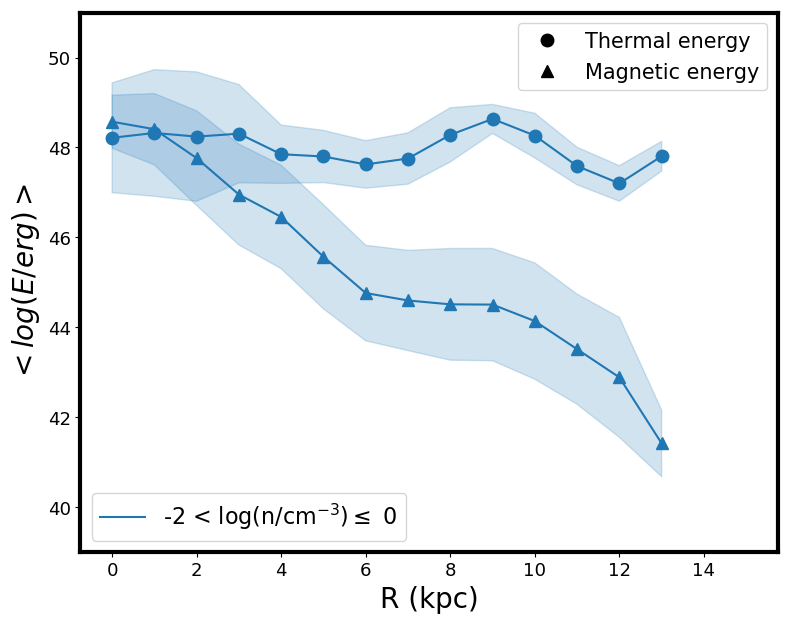}\\
     \rotatebox{90}{\hspace{1.6cm} Intermediate-density}
    \includegraphics[trim={0.cm 0.cm 0.cm 0.cm},clip,width=.4\linewidth]{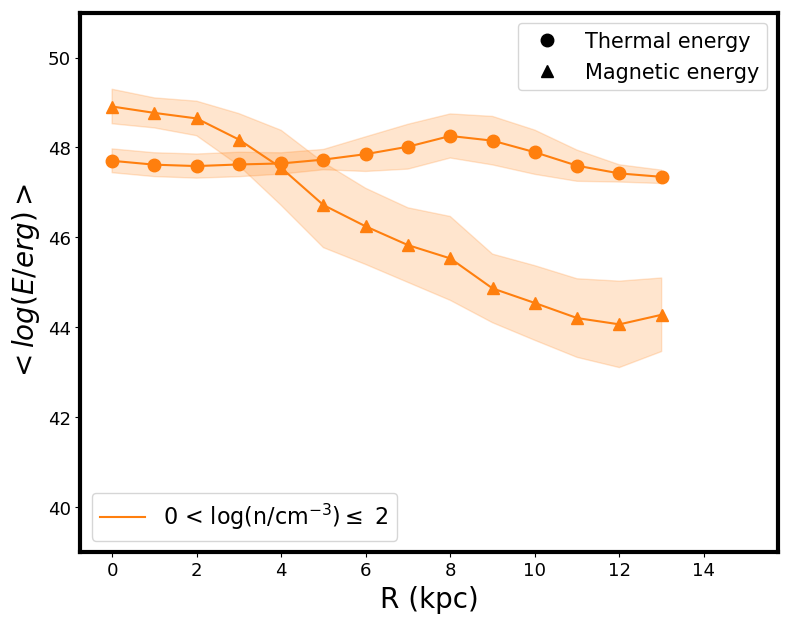} 
    \includegraphics[trim={0.cm 0.cm 0cm 0.cm},clip,width=.4\linewidth]{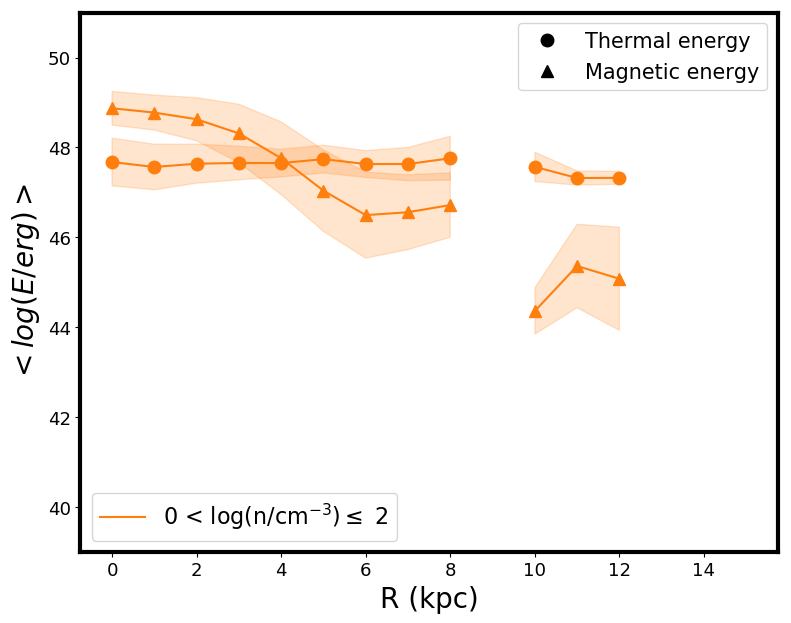}\\
    \rotatebox{90}{\hspace{2.15cm} High-density}
    \includegraphics[trim={0.cm 0.cm 0.cm 0.cm},clip,width=.4\linewidth]{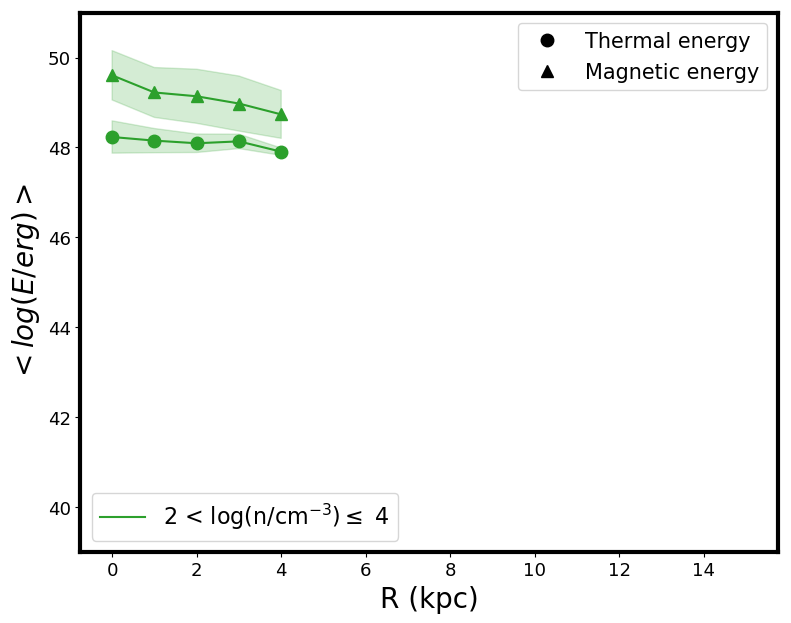} 
    \includegraphics[trim={0.cm 0.cm 0cm 0.cm},clip,width=.4\linewidth]{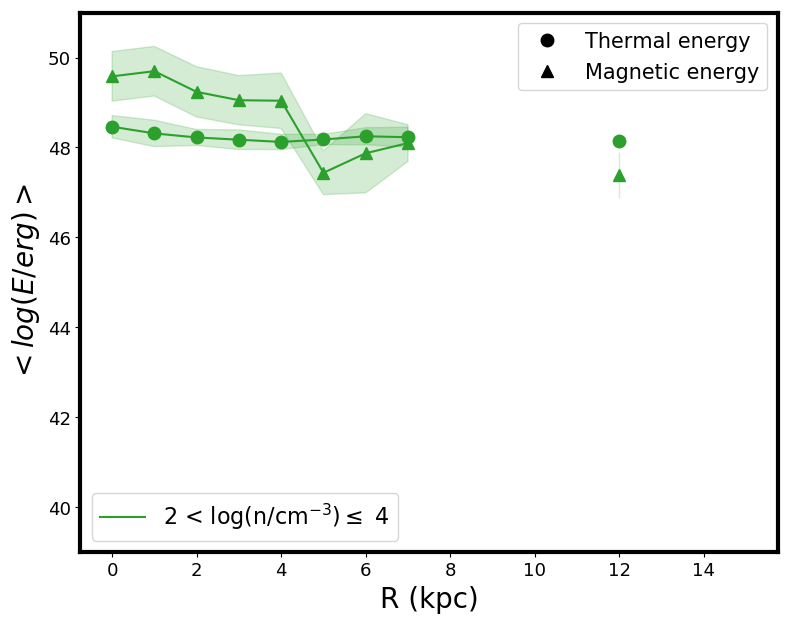}
    \caption{Radial profiles of the mean of the logarithm of thermal and magnetic energies for low, medium and high densities at 200 Myrs.}
    \label{fig:energy_profiles_200}
\end{figure*}

At 300 Myrs, the low-density gas is magnetically dominated for the first 1.5 kpc in both models, while beyond this region, it becomes thermally dominated. It is worth to note that the radius of model R extends to approximately 11 kpc, whereas model T reaches around 13.5 kpc.
The medium-density gas remains magnetically dominated for the initial 4 kpc from the center in both models, after which it transits to thermal dominance, consistent with its state at 200 Myrs.
The high-density gas extends to larger radii compared to the 200 Myrs snapshot in model T. In both models, they maintain magnetic dominance up to 4.5 kpc, after which they transit to thermal dominance.

\begin{figure*}[!ht]
    \centering
    t=300 Myrs
    \vspace{0.5cm}\\
    \hspace{1.0cm} Model T  \hspace{5.1cm} Model R \hspace{8.cm}\\
    \rotatebox{90}{\hspace{2.1cm} Low-density}
    \includegraphics[trim={0.cm 0.cm 0.cm 0.cm},clip,width=.4\linewidth]{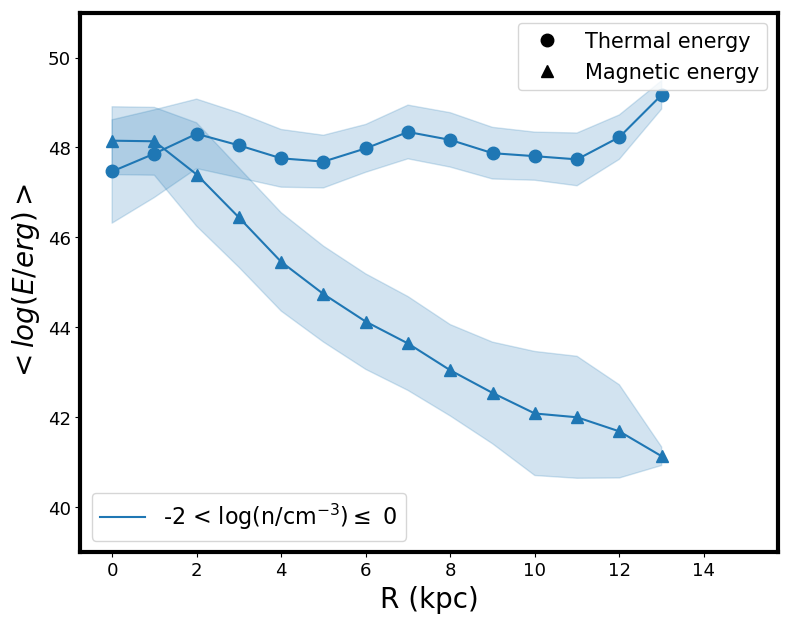} 
    \includegraphics[trim={0.cm 0.cm 0cm 0.cm},clip,width=.4\linewidth]{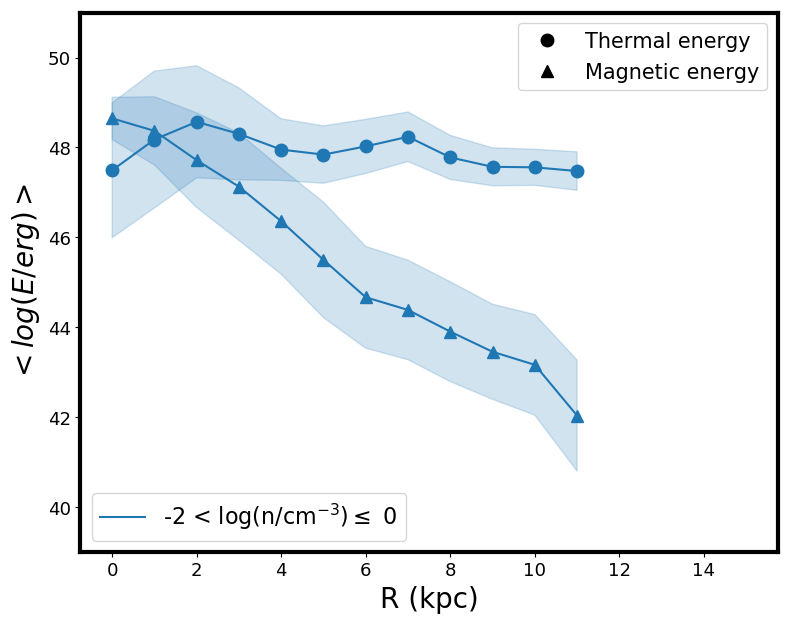}\\
     \rotatebox{90}{\hspace{1.6cm} Intermediate-density}
    \includegraphics[trim={0.cm 0.cm 0.cm 0.cm},clip,width=.4\linewidth]{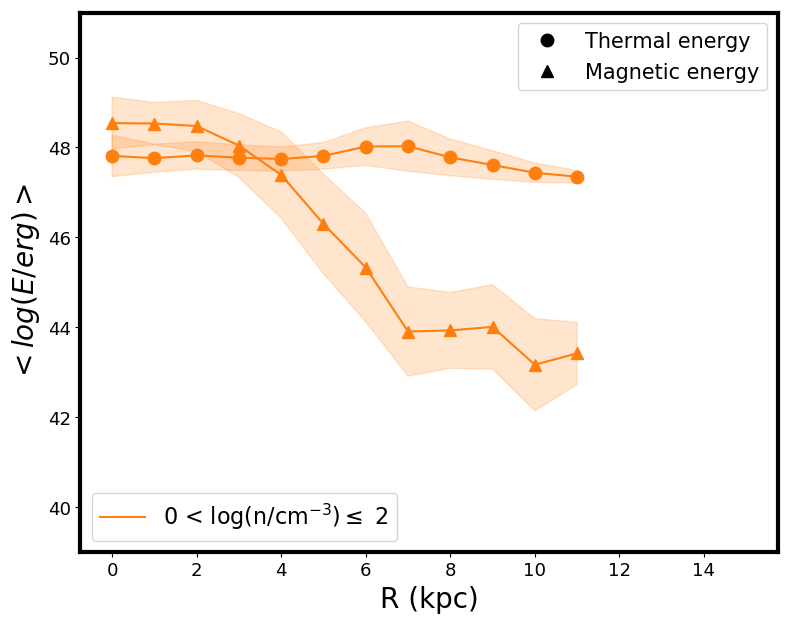} 
    \includegraphics[trim={0.cm 0.cm 0cm 0.cm},clip,width=.4\linewidth]{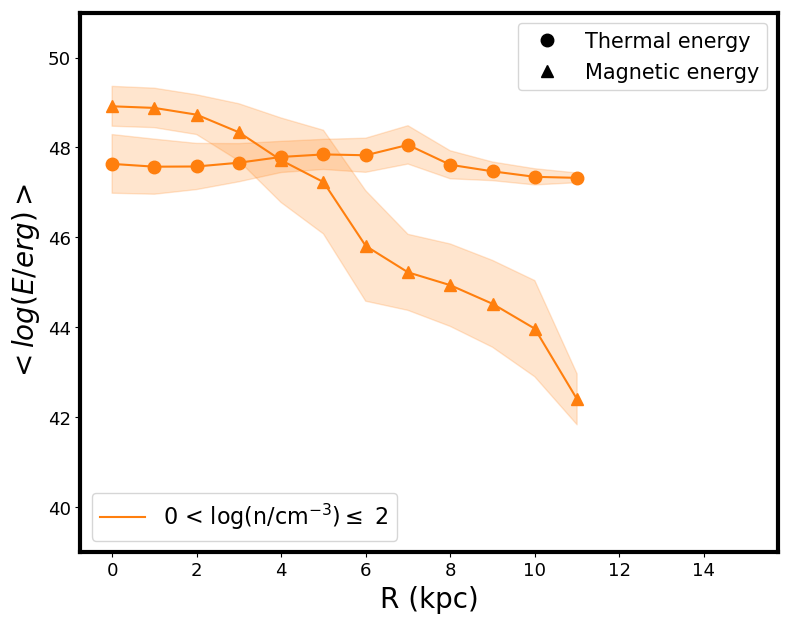}\\
    \rotatebox{90}{\hspace{2.15cm} High-density}
    \includegraphics[trim={0.cm 0.cm 0.cm 0.cm},clip,width=.4\linewidth]{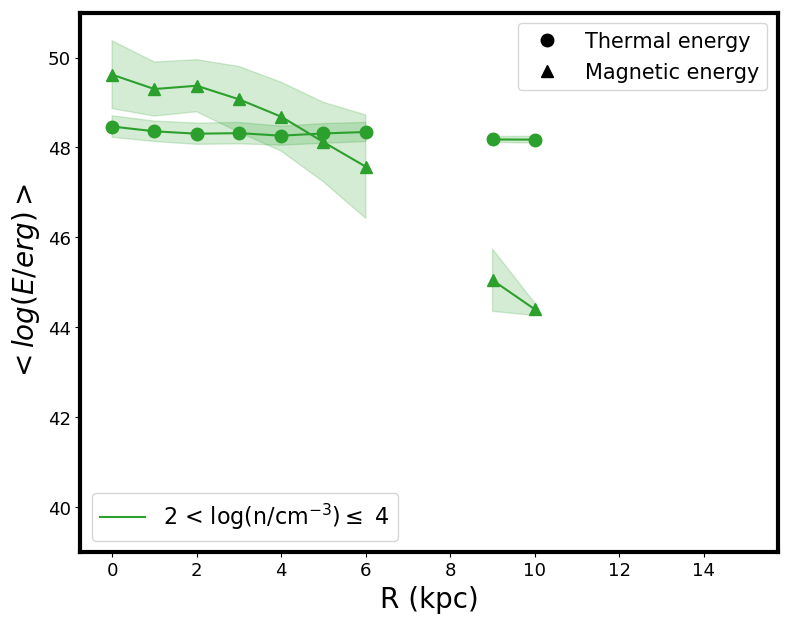} 
    \includegraphics[trim={0.cm 0.cm 0cm 0.cm},clip,width=.4\linewidth]{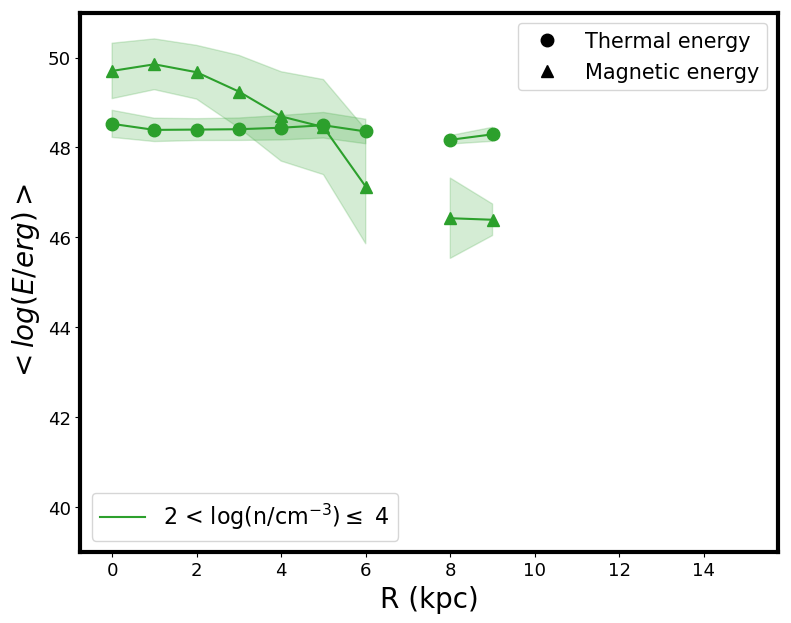}
    \caption{Radial profiles of the mean of the logarithm of thermal and magnetic energies for low, medium and high densities at 300 Myrs.}
    \label{fig:energy_profiles_300}
\end{figure*}

At 500 Myrs, there are notable differences in the behavior of low-density gas between the two models. In model T, there is a transition from equipartition in the central region to slight magnetic dominance extending up to 3 kpc. Beyond this point, thermal energy prevails. Conversely, in model R, the gas exhibits strong magnetic dominance from the center up to 3 kpc, with the magnetic energy being approximately two orders of magnitude higher than thermal energy. At 4 kpc, it reaches equipartition before transitioning to thermal dominance.
The medium-density gas maintains magnetic dominance for the initial 4 kpc from the center in model T and up to 5 kpc in model R, with model R having higher magnetic energy compared to model T in this region. Subsequently, both models transition to thermal dominance.
The high-density gas is magnetically dominated up to 4 kpc in model T and up to 6 kpc in model R.

Overall, the low-density gas is thermally dominated for the majority of the disk, but the central regions appear to be magnetically dominated, particularly in the case of model R, where the extent of magnetic dominance increases over time.
In the medium-density gas, both models consistently maintain magnetic dominance for the initial few kpc from the center before transitioning to thermal dominance. Notably, model R often exhibits higher magnetic energy compared to model T in this region.
For the high-density gas in both models, magnetic dominance is the initial state, followed by a transition to thermal dominance at certain radii.
In summary, these findings, in line with the patterns observed in the previous plots, suggest that model R consistently demonstrates a higher prevalence of magnetic dominance across various density regimes and time snapshots compared to model T.

\begin{figure*}[!ht]
    \centering
    t=500 Myrs
    \vspace{0.5cm}\\
    \hspace{1.0cm} Model T  \hspace{5.1cm} Model R \hspace{8.cm}\\
    \rotatebox{90}{\hspace{2.1cm} Low-density}
    \includegraphics[trim={0.cm 0.cm 0.cm 0.cm},clip,width=.4\linewidth]{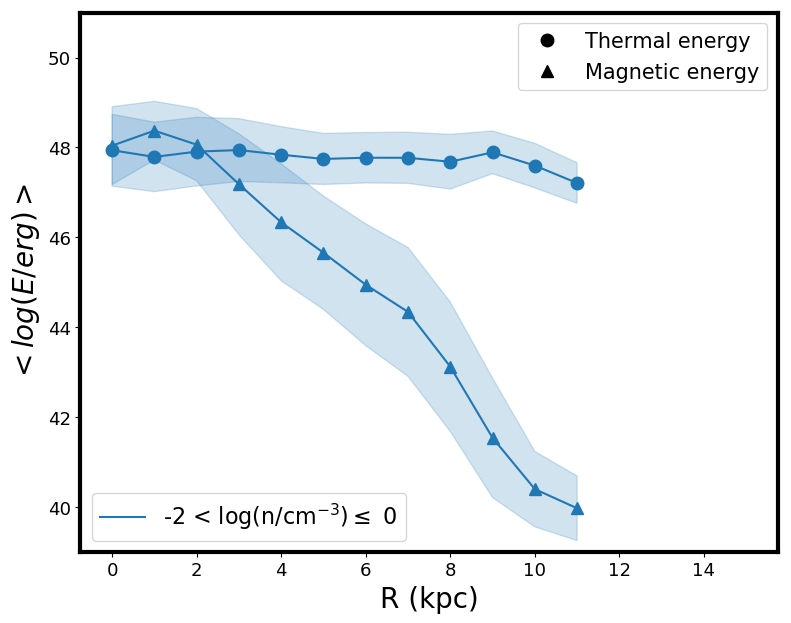} 
    \includegraphics[trim={0.cm 0.cm 0cm 0.cm},clip,width=.4\linewidth]{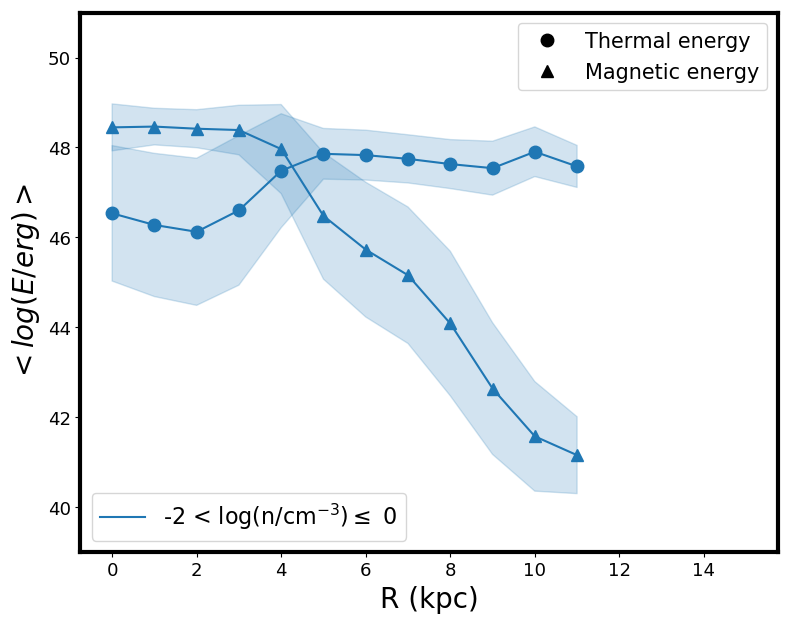}\\
     \rotatebox{90}{\hspace{1.6cm} Intermediate-density}
    \includegraphics[trim={0.cm 0.cm 0.cm 0.cm},clip,width=.4\linewidth]{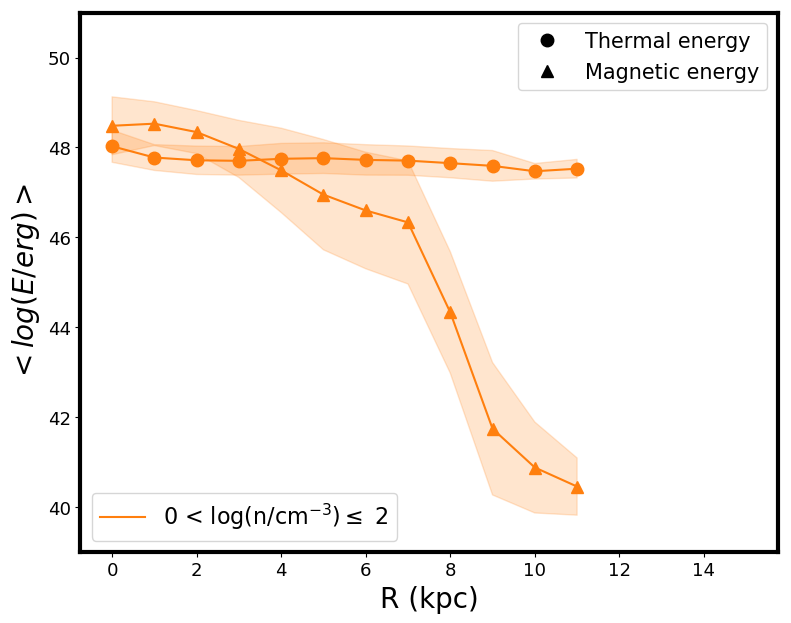} 
    \includegraphics[trim={0.cm 0.cm 0cm 0.cm},clip,width=.4\linewidth]{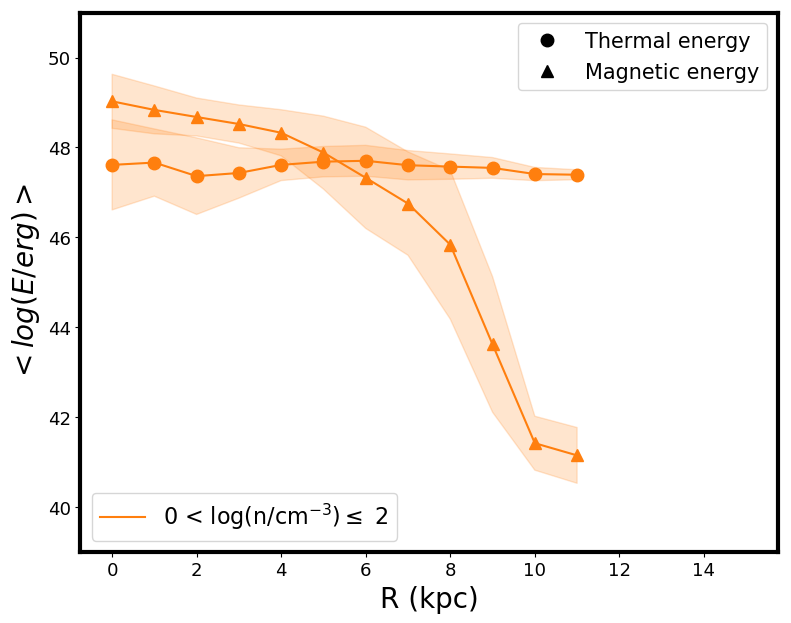}\\
    \rotatebox{90}{\hspace{2.15cm} High-density}
    \includegraphics[trim={0.cm 0.cm 0.cm 0.cm},clip,width=.4\linewidth]{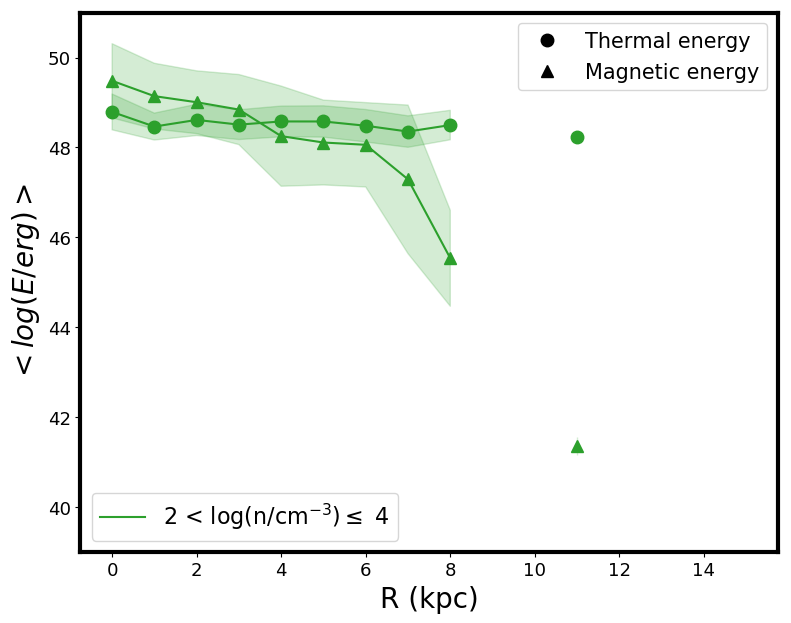} 
    \includegraphics[trim={0.cm 0.cm 0cm 0.cm},clip,width=.4\linewidth]{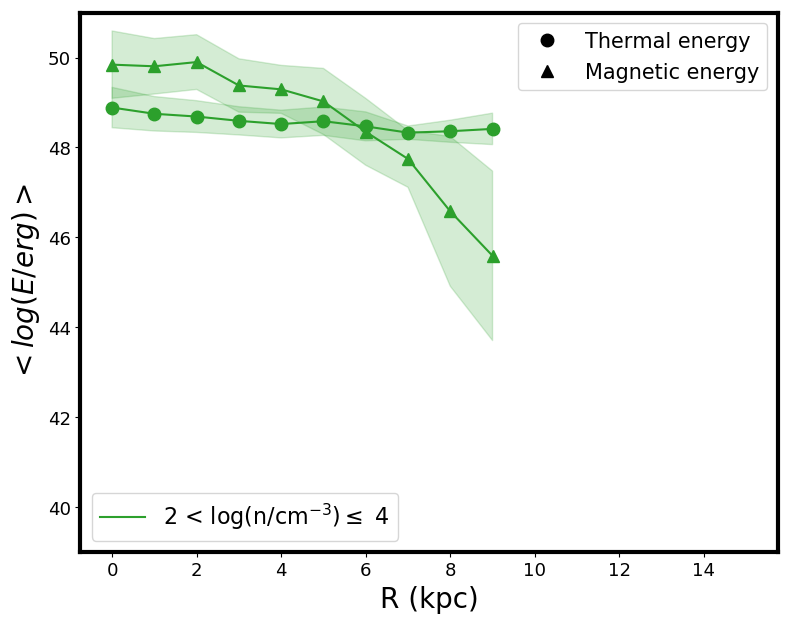}
    \caption{Radial profiles of the mean of the logarithm of thermal and magnetic energies for low, medium and high densities at 500 Myrs.}
    \label{fig:energy_profiles_500}
\end{figure*}

\end{appendix}

\end{document}